\documentclass[onecolumn]{IEEEtran}
\usepackage{amsmath}
\usepackage{mathrsfs}
\usepackage{accents}
\usepackage{setspace}
\usepackage{url}
\usepackage{algorithm}
\usepackage{amsfonts}
\usepackage{amssymb}
\usepackage{mathrsfs}
\usepackage{euscript}
\usepackage{graphicx}
\usepackage{multirow}
\usepackage{cite}
\usepackage[T1]{fontenc}
\usepackage{graphics}
\usepackage[protrusion=true,expansion=true]{microtype} 
\usepackage{wrapfig} 
\usepackage{epstopdf}
\usepackage{array}
\usepackage{booktabs}
\newtheorem{theorem}{Theorem}
\newtheorem{lemma}{Lemma}
\usepackage{tabularx}
\usepackage{tabulary}
\newtheorem{remark}{Remark}
\usepackage{subfig}
\newtheorem{proposition}{Proposition}
\usepackage[USenglish,UKenglish,french,spanish,italian]{babel}
\usepackage[nodayofweek,level]{datetime}
\newcommand{\mydate}{\formatdate{10}{05}{2017}}
\newcommand\blfootnote[1]{%
	\begingroup
	\renewcommand\thefootnote{}\footnote{#1}%
	\endgroup
}

\begin{document}

\begin{titlepage}

\begin{tabular}{l r}

\includegraphics[scale=0.3]{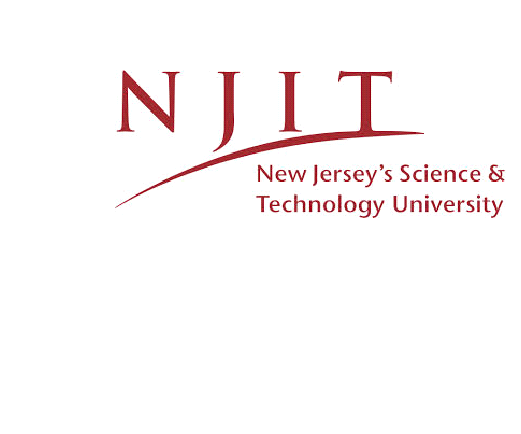} \hspace{10cm} & \includegraphics[scale=0.3]{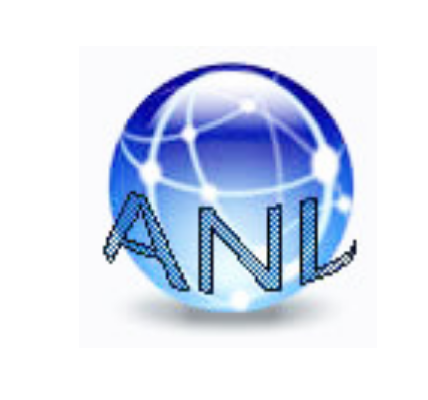}

\end{tabular}

\begin{center}

\textsc{\LARGE Joint Spectrum Allocation and Structure Optimization in Green Powered Heterogeneous Cognitive Radio Networks}\\[4cm]

{\Large \textsc{Ali Shahini}}\\ 
{\Large \textsc{Nirwan Ansari}}\\ 
[4cm]

{}
{\textsc{TR-ANL-2017-002}\\
\selectlanguage{USenglish}
\large \mydate} \\[4cm]

{\textsc{Advanced Networking Laboratory}}\\
{\textsc{Department of Electrical and Computer Engineering}}\\
{\textsc{New Jersy Institute of Technology}}\\[1.5cm]
\vfill

\end{center}

\end{titlepage}

\selectlanguage{USenglish}
\begin{spacing}{2}
\begin{abstract}
	We aim at maximizing the sum rate of secondary users (SUs) in OFDM-based Heterogeneous Cognitive Radio (CR) Networks using RF energy harvesting. Assuming SUs operate in a time switching fashion, each time slot is partitioned into three non-overlapping parts devoted for energy harvesting, spectrum sensing and data transmission. The general problem of joint resource allocation and structure optimization is formulated as a Mixed Integer Nonlinear Programming task which is NP-hard and intractable. Thus, we propose to tackle it by decomposing it into two subproblems. We first propose a sub-channel allocation scheme to approximately satisfy SUs' rate requirements and remove the integer constraints. For the second step, we prove that the general optimization problem is reduced to a convex optimization task. Considering the trade-off among fractions of each time slot, we focus on optimizing the time slot structures of SUs that maximize the total throughput while guaranteeing the rate requirements of both real-time and non-real-time SUs. Since the reduced optimization problem does not have a simple closed-form solution, we thus propose a near optimal closed-form solution by utilizing Lambert-W function. We also exploit iterative gradient method based on Lagrangian dual decomposition to achieve near optimal solutions. Simulation results are presented to validate the optimality of the proposed schemes.
\end{abstract}

\begin{IEEEkeywords}
	Resource Allocation, Energy Harvesting, Time Slot Optimization, Heterogeneous Cognitive Radio Network, Cooperative Spectrum Sensing.
\end{IEEEkeywords}
\blfootnote{The authors are with Advanced Networking Laboratory, the Helen and John C. Hartmann Department of Electrical and Computer Engineering, New Jersey Institute of Technology, Newark, NJ 07102.\protect~E-mail: {ali.shahini, nirwan.ansari}@njit.edu. Part of this work was presented at the Sarnoff Symposium, Newark, NJ, September 2016~\cite{00237}. This work was supported in part by NSF under grant no. CNS-1320468.}

\IEEEPARstart{T}{he} available radio frequency spectrum is getting crowded by the rapid growth of wireless applications and higher data rate devices~\cite{NirwanWileyBook,00117}. However, owing to inefficient conventional regulatory policies, a considerable amount of the radio spectrum is greatly underutilized. Cognitive Radio (CR) networks, as a promising paradigm with great potential of enhancing the spectrum utilization, allows efficient spectrum sharing between Primary Users (PUs) and Secondary Users (SUs)~\cite{00115}. In a CR network, SUs are allowed to sense the radio spectrum and occupy spectrum holes (i.e., spectral bands not utilized by PUs~\cite{00115}) in an opportunistic manner~\cite{00109-3}. The CR system has different functionalities in which spectrum sensing is considered to be the most challenging part of these systems~\cite{00120}. In practice, spectrum sensing cannot be reliably achieved by SUs due to shadowing and multipath fading. To alleviate the adverse impact of fading and achieve reliable spectrum sensing, cooperative spectrum sensing has been proposed and investigated~\cite{00120-16,00118,00120-18,00121,00126}. However, this functionality of sensing the radio spectrum incurs additional energy consumption.

Recent advances in energy harvesting are empowering the green powered CR network, in which SUs are equipped with energy harvesting capabilities to capture and store ambient energy which can significantly reduce carbon footprints~\cite{0001,00119,0001-14,0001-64}. In~\cite{0070-5}, the energy efficient resource allocation problem in heterogeneous CR systems is formulated and an iterative-based algorithm is proposed to solve the energy efficient resource allocation problem. Varshney ~\cite{0070-7} proposed a capacity-energy function and the idea of simultaneous data and energy transmission. Yin \textit{et al.}~\cite{00070} studied the duration of harvesting and number of sensed channels in one time slot. Their general problem is formulated as a mixed integer non-linear programming (MINLP) problem to maximize the achievable throughput of one SU with perfect spectrum sensing and without considering interference. However, in practical wireless systems, there are inevitable sensing errors stemmed from estimation errors, quantization errors and feedback delays. This imperfect spectrum sensing leads to substantial interference to the PUs caused by SUs. Thus, in order to prevent performance degradation of PUs, there should be a flexible physical layer for the CR system to control the interference generated by SUs.

Orthogonal Frequency Division Multiplexing (OFDM) is commonly known as a promising air interface for CR systems due to its great flexibility of radio resource allocation~\cite{00109-4}. In~\cite{00109}, sub-channel allocation and power allocation schemes have been incorporated in the OFDM based CR network with imperfect spectrum sensing. Wang \textit{et al.}~\cite{00109-17} proposed the sum capacity maximization for a CR system with a low complex algorithm while satisfying SUs' rate requirements. In~\cite{00122-20}, fair resource allocation has been proposed for CR and femtocell networks. However, imperfect spectrum sensing has not been considered in~\cite{00109-17,00122-20}. Since it is extremely difficult to attain perfect spectrum sensing in practical CR systems, sub-channel allocation with imperfect spectrum sensing should be considered. To the best of our knowledge, interference-aware resource allocation and structure optimization for energy harvesting enabled SUs for OFDM based heterogeneous CR networks with imperfect cooperative spectrum sensing has not been studied.

In this paper, we investigate the joint sub-channel allocation and structure optimization for OFDM based heterogeneous CR networks by using RF energy harvesting, with consideration of interference limitations, imperfect spectrum sensing, and various rate requirements of SUs. Since SUs are assumed to operate in a time switching fashion, each time slot is partitioned into three non-overlapping fractions devoted for energy harvesting, spectrum sensing and data transmission. The first part of each time slot is allocated for energy harvesting characterized by a metric called harvesting ratio. Although the higher harvesting ratio implies more time allocated for energy harvesting (extracting more energy), it leads to less remaining time for data transmission. Hence, the ultimate goals are to find optimal harvesting ratios (best tradeoff between operations) of SUs and optimal sub-channel allocation to SUs in order to maximize the total throughput of the CR network. The main contributions of this paper can be summarized as follows.
\begin{itemize}
	\item We formulate the joint sub-channel allocation and structure optimization as a sum rate maximization of SUs in OFDM based heterogeneous CR networks by using RF energy harvesting, where interference limits are imposed to protect the PUs, rate requirements for both real-time and non-real-time SUs are considered to guarantee fairness for SUs in each CR network, and cooperative spectrum sensing is employed to provide more reliable results of channel sensing while considering imperfect spectrum sensing. 
	\item We analyze the general optimization problem and show that it is MINLP, computationally intractable and NP-hard. Thus, we propose to address the general problem in two steps by mathematically decomposing it into two subproblems. We thus propose a sub-channel allocation scheme based on a factor called Energy Figure of Merit to approximately satisfy SUs' rate requirements and remove the integer constraints. In the sub-channel allocation process,	the real-time (RT) SUs have higher priority to receive sub-channels as compared to non-real-time (NRT) SUs.
	\item We prove that the general optimization problem is reduced to a nonlinear convex optimization task. Since the reduced optimization problem does not have a simple closed-form solution for optimal harvesting ratios of SUs, we thus propose a near optimal closed-form solution by utilizing Lambert-W function to obtain optimal harvesting ratios. In order to derive the closed-form solution, we prove a lemma (Lemma 3) that can be utilized for other similar problems. We also exploit the iterative gradient method based on Lagrangian dual decomposition to achieve near optimal solutions.
	\item The proposed methods and algorithms are evaluated by extensive simulations. The simulation results show that the proposed sub-channel allocation scheme outperforms the existing schemes especially when the number of available sub-channels are low. The simulation and numerical results verify the effectiveness of our closed-form solution for harvesting ratios of SUs, where the performance gap from the optimal solution is less than 3.5$\%$ for various cases. Further, we analyze the performance of our system in terms of interference protection of PUs, and different SUs' required rate constraints.	
\end{itemize}
\begin{figure}[t]
	\centering
	\includegraphics[width=9cm,height=9cm]{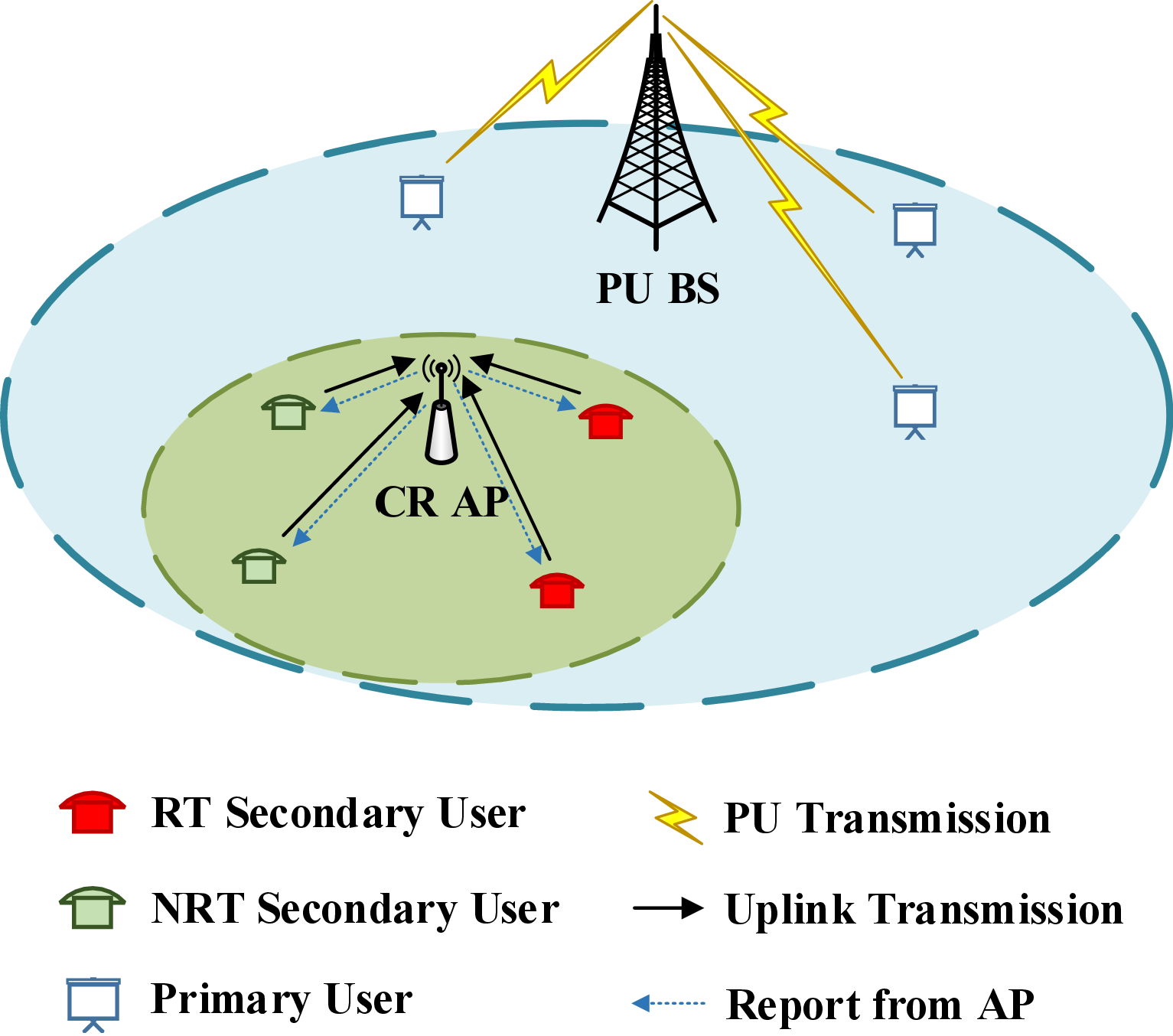}
	\caption{System model of the heterogeneous CR network. Both RT SUs and NRT SUs are shown around one AP.}
	\label{fig:SystemModelFigure}
\end{figure}
The remainder of this paper is organized as follows. In Section \ref{sec:System Model}, we first describe the system model, cooperative spectrum sensing and time slot structure. In Section \ref{ProbFormul}, we formulate the general sum rate maximization problem. In Section \ref{Sub-Alloc}, we discuss our solution methodology, propose a sub-channel allocation scheme and optimize the time slot structures of SUs. In Section \ref{sec:simultion}, numerical results and simulations are presented with analysis on sub-channel allocation and optimal time slot structure. Finally, we conclude the paper in Section \ref{sec:conclusion}.
\section{System Model}\label{sec:System Model}
Consider an uplink OFDM-based heterogeneous CR network compromising $L$ PUs denoted by ${\cal L} = \{ 1,2,...,L\}$ and $K$ self-powered SUs represented by ${\cal K} = \{ 1,2,...,K\}$ with $N$ OFDM licensed sub-channels operating in the slotted mode. These aggregated OFDM sub-channels constitute the licensed spectrum such that parts of the spectrum are registered by PUs. The SUs harvest energy from ambient radio signals and have no other power supplies. To support diverse services, the CR network has $i_{0}$ NRT SUs with rate constraints $\zeta_i$, and $K-i_{0}$ RT SUs with minimum required rate $R_i^{req}$. In other words, the NRT SUs is denoted by subset ${{\cal K}_N} = \{ 1,...,{i_0}\}$ and the subset ${{\cal K}_R} = \{ {i_0} + 1,...,K\} $ represents the RT SUs. The licensed sub-channels are opportunistically utilized by SUs via an Access Point (AP). We assume that the SUs have perfect knowledge of Channel State Information (CSI) between their transmitters and the AP receiver. In our work, the general system model of a heterogeneous CR network is illustrated in Fig.~\ref{fig:SystemModelFigure}.
\begin{figure}[t]
	\centering
	\includegraphics[width=10cm,height=6cm]{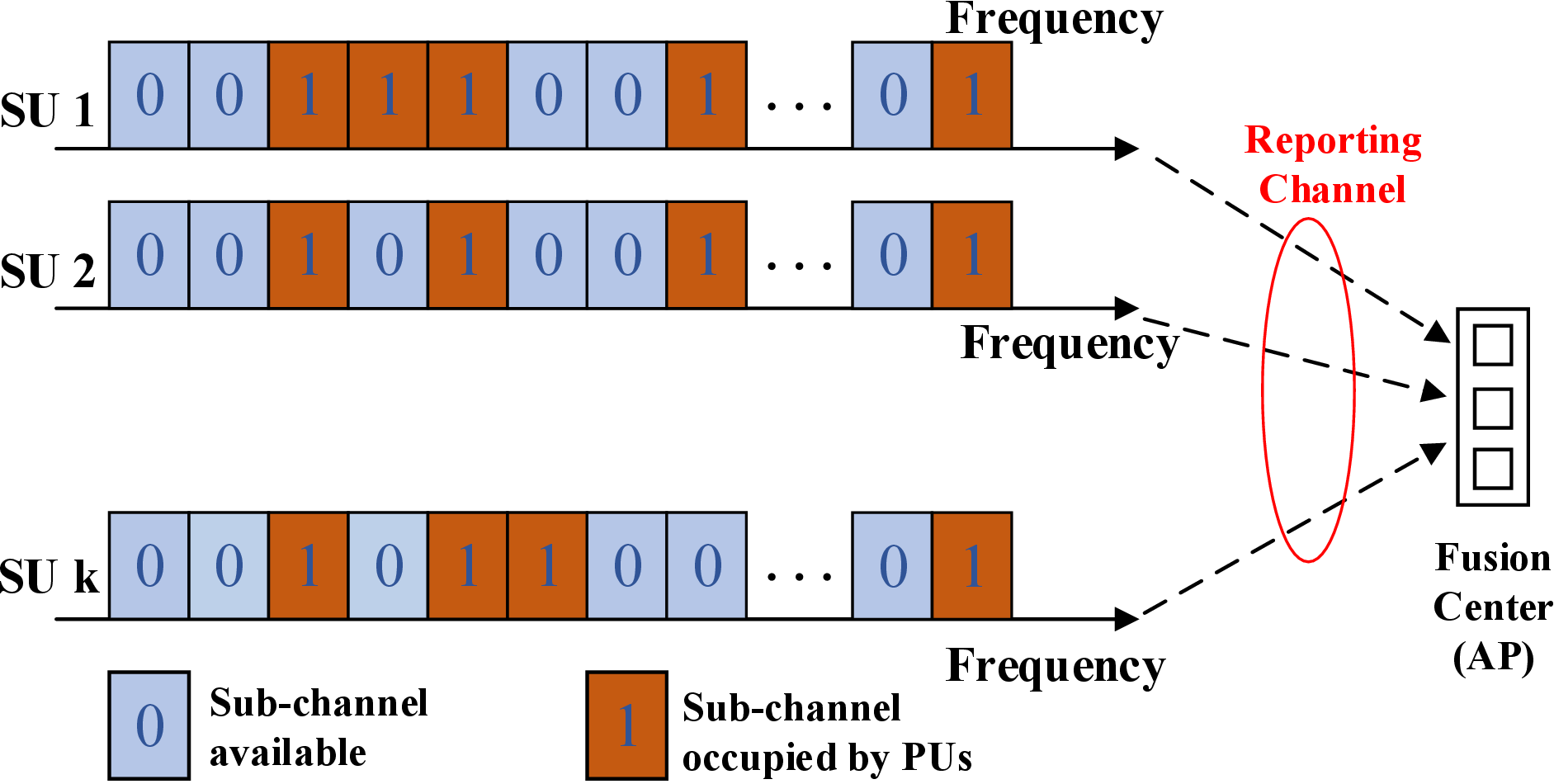}
	\caption{The SUs report their spectrum sensing results to the FC for making the final decision.}
	\label{fig:SpecHolesReports}
\end{figure}

\subsection{Cooperative Spectrum Sensing}
In our system model, each SU does a local spectrum sensing concerning the presence or absence of PUs. It is assumed that the sensing results of SUs are independent and SUs sense all the PUs' sub-channels appointed by the AP. In order to reduce the spectrum sensing errors arisen from fading and shadowing, Cooperative Spectrum Sensing (CSS) has been exploited. In our CSS scenario, multiple SUs sense the licensed sub-channels independently, and the PUs' activities can be predicted by the AP \cite{00145} using the collected sensing results of SUs. Fig.~\ref{fig:SpecHolesReports} illustrates a CSS scenario in which $K$ SUs independently sense $N$ sub-channels and identify the absence and presence of PUs by $0$ and $1$ bits, respectively. In fact, these one-bit decisions are reported to a Fusion Center (FC) which is located in the AP. Then, FC applies a fusion strategy and generates final decisions regarding availability of OFDM licensed sub-channels. In this work, we assume that each SU applies the Energy Detection (ED) strategy which has low computational complexities \cite{00121}, and the FC follows the Majority rule (generalized as $k$-out-of-$n$) \cite{00120}. Finally, the available sub-channels of a subset ${\cal M} = \{ 1,2,...,M\}$ among all the licensed sub-channels of a subset ${\cal N} = \{ 1,2,...,N\}$ are identified by the AP and replied to SUs at the beginning of each time slot.
\subsection{Time Slot Model}

In this work, an OFDM based CR system with SUs operating in a slotted mode is considered. Each SU in one time slot, is expected to do the following operations: (1) energy harvesting, (2) contributing in cooperative spectrum sensing, and (3) data transmission. In each time slot with duration $T$, due to the duplex-constrained hardware~\cite{00123}, the energy harvesting process and energy consuming process for SUs should be scheduled in a time switching fashion~\cite{0001}. Thus, we assume SUs operate in a time switching fashion, and the time slot is partitioned into three non-overlapping parts devoted for energy harvesting, spectrum sensing and data transmission, respectively. Hence, the first fraction of each time slot (harvesting ratio: $\theta_i, ~\forall i \in {\cal K}$) is allocated to energy harvesting process. Although traditional energy-constrained wireless networks are powered by fixed energy sources like batteries, it may be expensive, inconvenient\footnote{One of the dominant barriers to implementing IoT networks is providing adequate energy for operating the network in a self-sufficient manner \cite{00215}.}, and even hazardous\footnote{Battery replacements can be dangerous in a toxic environment \cite{00216}.} \cite{00216}. Thus, the SUs are considered to have no power supplies other than harvesting energy from ambient radio signals\footnote{Energy-harvesting circuits (e.g., P2110B Powercast receiver \cite{Powercast}) can harvest micro-watts to milliwatts of power within the range of several meters for a transmit power of 1 W and a carrier frequency of 915MHz \cite{Powercast}.}. Then, spectrum sensing, which depends on SUs' location and performance of sensing, can be accomplished in the second step of each time slot. During the sensing time ($\tau_{s_i},~\forall i \in {\cal K}$), SUs sense the licensed sub-channels and report their local sensing results to the FC, where the final decision regarding availability of sub-channels would be finalized. The third part of the time slot is utilized for data transmission. In fact, the available sub-channels are allocated to SUs by AP at the beginning of each time slot and SUs transmits data using all the remaining harvested energy after the spectrum sensing phase.
\begin{figure*}[t]
	\centering
	\includegraphics[width=18cm,height=8cm]{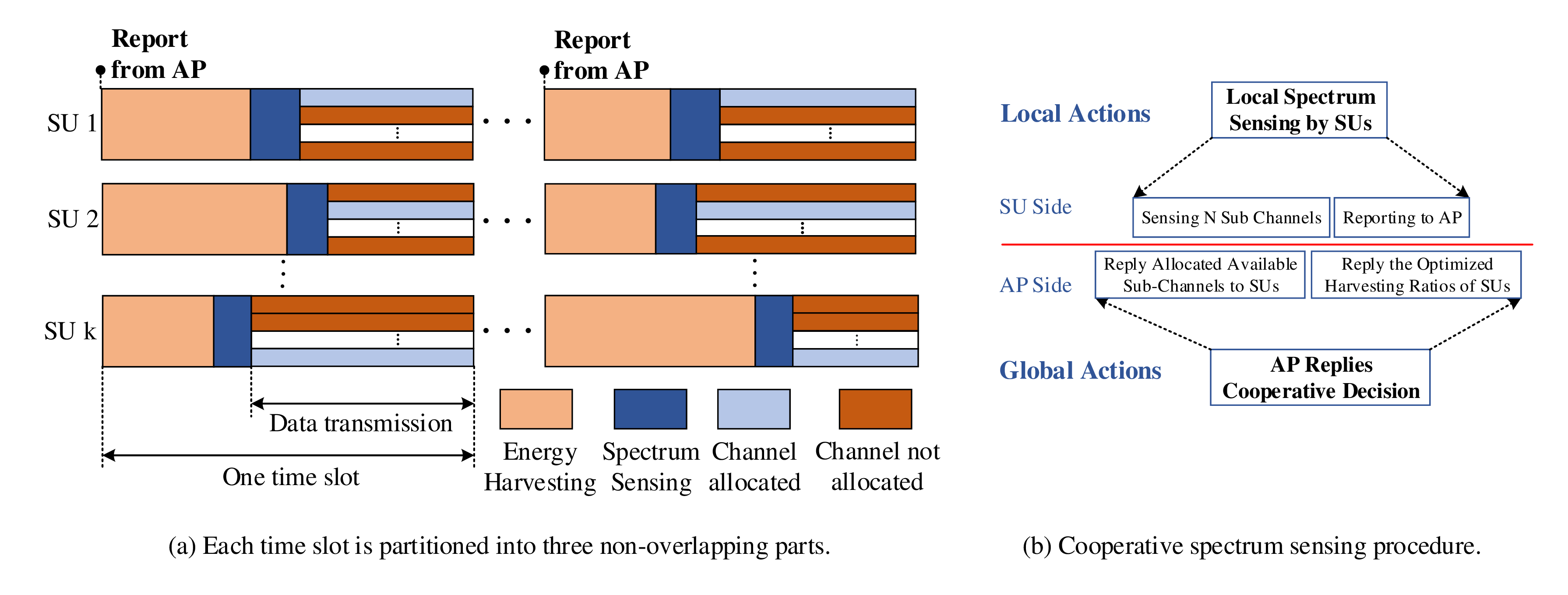}
	\caption{The time slot structure with energy harvesting for multiple SUs.}
	\label{fig:TimeSlotFigure}
\end{figure*}
The time slot structures for $K$ SUs are illustrated in Fig.~\ref{fig:TimeSlotFigure}(a), such that each SU has different harvesting ratio, sensing time and transmission time. At the beginning of each time slot, SUs receive reports from AP regarding the SUs' sub-channel allocations and optimized harvesting ratios. Hence, SUs start extracting energy from ambient radio signals during interval (0, $\theta_{i}T$] and store it in a storage for future use within the time slot only. Then, SUs switch from harvesting to spectrum sensing during ($\theta_{i}T$, $\theta_{i}T$+${\tau_{s_{i}}}$]. Note that, SUs have different performance of sensing, and thus various sensing time ${\tau_{s_{i}}}$. Meanwhile, AP receives the spectrum sensing results, make the final decision, and report it to SUs at the beginning of next time slot. Furthermore, during the third fractions of the time slot, SUs start transmitting data using the harvested energy.

Fig.~\ref{fig:TimeSlotFigure}(b), describes the cooperative spectrum sensing procedure, where the SUs operate and report local channel sensing to the AP during the sensing time in each time slot. Hence, cooperative decision strategies are utilized at the AP side to make reliable sensing results and report them to SUs for the next time slot. However, perfect spectrum sensing in practical CR systems cannot be accomplished due to imperfect channel sensing with typical sensing errors. As a matter of fact, spectrum sensing errors are generally categorized into two groups: \emph{miss-detections} and \emph{false alarms}. Miss-detection happens when the CR network fails to detect the PU signals and false alarm occurs when the CR system identifies an actually vacant sub-band as being used by the PU. Clearly, co-channel interferences to the PUs arise from miss-detection errors and spectrum efficiency utilization is degraded by false alarm errors. Throughout this work, $Q_\ell ^m$ and $Q_\ell ^f$ denote the probabilities of miss-detection and false alarm on the $\ell^{th}$ sub-channel, respectively.

\begin{table}  \label{tableHO}	\caption{Spectrum Sensing Results of SUs}
	\centering
	\begin{tabular}{||l l l l ||} 
		\hline
		No.&Actual&Sensing&Probabilities\\ [0.5ex] 
		\hline\hline
		1 & ${\cal H}_{1,\ell }$ & ${\cal S}_{1,\ell }$ &${P_d}=P\{ {{\cal S}_{1,\ell }}|{{\cal H}_{1,\ell }}\}  = 1 - Q_\ell ^m$\\ 
		\hline
		2 & ${\cal H}_{1,\ell }$ & ${\cal S}_{0,\ell }$ &${P_m}=P\{ {{\cal S}_{0,\ell }}|{{\cal H}_{1,\ell }}\}  = Q_\ell ^m$ \\ 
		\hline
		3 & ${\cal H}_{0,\ell }$ & ${\cal S}_{1,\ell }$ &${P_f}=P\{ {{\cal S}_{1,\ell }}|{{\cal H}_{0,\ell }}\}  = Q_\ell ^f$ \\ 
		\hline
		4 & ${\cal H}_{0,\ell }$ & ${\cal S}_{0,\ell }$ &$P\{ {{\cal S}_{0,\ell }}|{{\cal H}_{0,\ell }}\}  = 1 - Q_\ell ^f$ \\ 
		\hline
	\end{tabular}
\end{table}
\begin{table}  \label{tableSymbols}	\caption{List of Symbol Notations and Description}
	\centering
	\begin{tabular}{ |l|p{6.5cm}| }
		\hline
		\textbf{Symbols} & \textbf{Descriptions}\\\hline \hline
		L ($\mathcal{L}$)& The total number of PUs (the set of PUs ) \\\hline
		K ($\mathcal{K}$)& The total number of SUs (the set of SUs) \\\hline
		$\mathcal{K}_R$ ($\mathcal{K}_N$) & The set of real time SUs (the set of non real time SUs) \\\hline
		M & The number of available sub-channels determined by FC\\\hline
		$\mathcal{M}$ & The set of available sub-channel determined by FC\\\hline
		N & The total number of sub-channels\\\hline
		T & The duration of time slot\\\hline
		${{f_{s}}}$   & The starting frequency \\\hline
		$\omega $ & The bandwidth of each sub-channel\\\hline
		$t$ & The OFDM symbol duration \\\hline
		${{h_{i,j}}}$ & The channel gain between the $i^{th}$ SU and the AP over sub-channel $j$\\\hline
		$\Gamma $ & The SNR gap associated with BER\\\hline
		$\varphi (f)$ & The PSD of OFDM signal \\\hline
		${\mathcal{H}_{1,\ell }}$ & The presence of the PUs' signal on the $\ell^{th}$ sub-channel \\\hline
		${\mathcal{H}_{0,\ell }}$ & The absence of the PUs' signal on the $\ell^{th}$ sub-channel \\\hline
		${\mathcal{S}_{1,\ell }}$ & The $\ell^{th}$ sub-channel is determined available by the FC\\\hline
		${\mathcal{S}_{0,\ell }}$ & The $\ell^{th}$ sub-channel is determined unavailable by the FC\\\hline
		$Q_\ell ^m$ & The probability of mis-detection of $\ell^{th}$ sub-channel\\\hline
		$Q_\ell ^f$ & The probability of false alarm of $\ell^{th}$ sub-channel\\\hline
		${{\theta _i}}$ & The harvesting ratio of $i^{th}$ SU in each time slot \\\hline
		${{\tau _{{s_i}}}}$ & The sensing time of $i^{th}$ SU\\\hline
		${{\chi _i}}$ & The rate of energy harvesting for $i^{th}$ SU\\\hline
		${{\epsilon_{{s_i}}}}$ & The energy consumed by $i^{th}$ SU for sensing\\\hline
		${{I_{i,j,m}}}$ & The total interference introduced to the $m^{th}$ PU by the transmission of $i^{th}$ SU over $j^{th}$ sub-channel\\\hline
		${I_m^{th}}$ & The interference threshold of $m^{th}$ PU\\\hline
		${R_i^{req}}$ & The required rate of the $i^{th}$ SU for RT users\\\hline
		${{\zeta_i}}$ & The rate constraint of the $i^{th}$ SU for NRT users \\\hline
		${{f_{i,j}}}$ & The indicator function for assigning $j^{th}$ sub-channel to the $i^{th}$ SU\\\hline
		${{r_{i,j}}}$ & The transmission rate of the $i^{th}$ SU over $j^{th}$ sub-channel\\\hline
		$\alpha _{EFM}^i$ & The energy figure of merit factor for $i^{th}$ SU\\\hline
		${{\mathcal{D}_i}}$ & The set of assigned sub-channels to $i^{th}$ SU\\
		\hline
	\end{tabular}
\end{table}
Table 1 illustrates possible outcomes of spectrum sensing by SUs. Presence and absence of PUs can be represented by ${{\cal H}_{1,\ell }}$ and ${{\cal H}_{0,\ell }}$, while the sensing results of the $\ell^{th}$ sub-channel for availability and unavailability of PUs are denoted by ${{\cal S}_{1,\ell }}$ and ${{\cal S}_{0,\ell }}$, respectively. Moreover, ${P_d}$, ${P_m}$, and ${P_f}$ are probabilities of detection, miss-detection, and false alarm, respectively. The final decision regarding availability of licensed sub-channels is made by the FC at the AP, based on the sensed information of SUs. Meanwhile, one should consider the analyzing, processing, and optimization time of the AP as well as replying time of the final decision from the AP to SUs. Hence, considering this elapsed time by the AP, the final result of cooperative spectrum sensing regarding availability of sub-channels is known for SUs at the beginning of the next time slot. In fact, SUs receive reports from AP at the beginning of the time slot concerning the SUs' sub-channel allocations and optimized harvesting ratios. Note that we assume states of sub-channels do not change within a time slot and SUs have plenty of data in their buffers. The available sub-channels in the sub-band of the $m^{th}$ PU are denoted by subset ${\cal M}_{A,m}$, while the unavailable sub-channels are represented by subset ${\cal M}_{U,m}$. Some of frequently used notations and terminologies are summarized in Table 2.

\section{Problem Formulation} \label{ProbFormul}
In this part, an optimization framework is formulated to maximize the total throughput of a green powered OFDM based heterogeneous CR network under some practical considerations. In fact, the problem is a joint sub-channel allocation and structure optimization problem which aims at maximizing the SUs sum rate by allocating the optimal number of sub-channels to SUs and finding the optimal trade-off between fractions of the SUs' time slot.

Denote $\omega$ as the bandwidth of each OFDM sub-channel, and the range of nominal spectrum for the $\ell ^{th}$ sub-channel is from $f_{s}$+$(\ell -1)\omega$ to $f_{s}$+$\ell\omega$ ($f_{s}$ is the starting frequency). The amount of interference introduced to the $j^{th}$ sub-channel in the sub-band of the $m^{th}$ PU caused by the $i^{th}$ SU transmission over the $\ell$ sub-channel with unit transmission power can be expressed as~\cite{jsac28}
\begin{eqnarray}\label{Interference}
I_{i,j,m}^\ell  = \int_{(j - 1)\omega  - (\ell  - {\raise0.7ex\hbox{$1$} \!\mathord{\left/
			{\vphantom {1 2}}\right.\kern-\nulldelimiterspace}
		\!\lower0.7ex\hbox{$2$}})\omega }^{j\omega  - (\ell  - {\raise0.7ex\hbox{$1$} \!\mathord{\left/
			{\vphantom {1 2}}\right.\kern-\nulldelimiterspace}
		\!\lower0.7ex\hbox{$2$}})\omega } {\varphi (f){g_{i,\ell ,m}}df},
\end{eqnarray}
where $\varphi (f) = t{(\frac{{\sin (\pi ft)}}{{\pi ft}})^2}$ represents the power spectrum density (PSD) of the OFDM signal ($t$ is the OFDM symbol duration). $g_{i,\ell ,m}$ denotes the power gain from the $i^{th}$ SU to the receiver of the $m^{th}$ PU on the $\ell$ sub-channel.

The probability of the CR system to make a correct decision that the $\ell^{th}$ sub-channel ($\ell  \in {\cal M}$) is truly used by a PU is denoted by $P_\ell ^1$:
\begin{equation} \label{P1L}
\begin{split}
P_\ell ^1 =&P\{ {{\cal H}_{1,\ell }}|{{\cal S}_{1,\ell }}\} \\=
&\frac{{P\{ {{\cal H}_{1,\ell }}\} P\{ {{\cal S}_{1,\ell }}|{{\cal H}_{1,\ell }}\}}}{{P\{ {{\cal H}_{1,\ell }}\} P\{ {{\cal S}_{1,\ell }}|{{\cal H}_{1,\ell }}\} + P\{ {{\cal H}_{0,\ell }}\} P\{ {{\cal S}_{1,\ell }}|{{\cal H}_{0,\ell }}\} }} \\=
& \frac{{Q_\ell ^L (1 - Q_\ell ^m)}}{{Q_\ell ^L (1 - Q_\ell ^m) + (1 - Q_\ell ^L)Q_\ell ^f}}.
\end{split}
\end{equation}
Likewise, $P_{\ell} ^2$ denotes the probability of the CR system making a decision that the $\ell^{th}$ sub-channel is available but it is truly occupied:
\begin{eqnarray} \label{P2L}
\begin{split}
P_\ell ^2 &= P\{ {{\cal H}_{1,\ell }}|{{\cal S}_{0,\ell }}\} \\
&= \frac{{P\{ {{\cal H}_{1,\ell }}\} P\{ {{\cal S}_{0,\ell }}|{{\cal H}_{1,\ell }}\} }}{{P\{ {{\cal H}_{1,\ell }}\} P\{ {{\cal S}_{0,\ell }}|{{\cal H}_{1,\ell }}\} + P\{ {{\cal H}_{0,\ell }}\} P\{ {{\cal S}_{0,\ell }}|{{\cal H}_{0,\ell }}\} }} \\
&= \frac{{Q_\ell ^L Q_\ell ^m}}{{Q_\ell ^L Q_\ell ^m + (1 - Q_\ell ^L)(1 - Q_\ell ^f)}},
\end{split}
\end{eqnarray}
where $Q_\ell ^L$ represents the \textit{a priori} probability that the sub-band of the $\ell^{th}$ sub-channel is used by PUs. Hence, the total interference introduced to the $m^{th}$ PU stemmed from the access of the $i^{th}$ SU on the $\ell^{th}$  sub-channel with unit transmission power is given as
\begin{eqnarray} \label{Imelli}
{I_{i,\ell ,m}} = \sum\limits_{j \in {{\cal M}_{A,m }}} {P_j^1I_{i,j,m}^\ell }  + \sum\limits_{j \in {{\cal M}_{U,m }}} {P_j^2I_{i,j,m}^\ell }.
\end{eqnarray}
Meanwhile, the rate of transmission of the $i^{th}$ SU over the sub-channel $j$ in one time slot can be expressed as
\begin{equation} \label{rij}
{r_{i,j}} = (1 - {\theta _i} - \frac{{{\tau _{{s_i}}}}}{T})
lo{g_2}(1 + \frac{{{{\left| {{h_{i,j}}} \right|}^2}({\chi_i}{\theta _i}T - {\epsilon_{{s_i}}})}}{{\Gamma (\omega{N_0} + I_{i})(T - {\theta _i}T - {\tau _{{s_i}}})}}),
\end{equation}
where $\theta_i$ denotes the harvesting ratio of the $i^{th}$ SU, $\chi_{i}$ is the energy harvesting rate of the $i^{th}$ SU, $h_{i,j}$ denotes the channel gain of the $i^{th}$ SU over sub-channel $j$, $N_{0}$ represents the additive white Gaussian noise, $\epsilon_{s_{i}}$ denotes the energy of sensing by the $i^{th}$ SU, and $\Gamma$ is the SNR gap. $\Gamma$ is associated with the bit-error-rate (BER) of un-coded MQAM and $\Gamma= -ln (5 \times BER) /1.5$ \cite{00109-31}. The interference introduced to the $i^{th}$ SU caused by the PUs' signals represented by $I_{i}$ is considered as noise and can be computed by the proposed method in \cite{00109-12}. Thus, the total transmission rate of the $i^{th}$ SU can be given as
\begin{equation} \label{RateUser}
{R_i} = \sum\limits_{j = 1}^M {{f_{i,j}}(1 - {\theta _i} - \frac{{{\tau _{{s_i}}}}}{T})lo{g_2}(1 + {H_{i,j}}\frac{{{\chi_i}{\theta _i}T - {\epsilon_{{s_i}}}}}{{T - {\theta _i}T - {\tau _{{s_i}}}}}).}
\end{equation}
The binary variable ${f_{i,j}} \in \{ 0,1\}$ is utilized to represent the sub-channel assignment between SU $i$ and sub-channel $j$:
\begin{eqnarray}
{f_{i,j}} = \left\{ {\begin{array}{*{20}{c}}
	{1,}& $if$~$sub-channel$~j~$is$~$assigned$~$to$~$SU$~i \\ 
	{0,}& $otherwise.$ 
	\end{array}} \right.
\end{eqnarray}
Note that the term of ${H_{i,j}} = \frac{{{{\left| {{h_{i,j}}} \right|}^2}}}{{\Gamma (\omega{N_0} + I_i)}}$ is used in Eq.~(\ref{RateUser}) for simplicity.

Finally, the general problem of the uplink sum rate maximization of SUs can be formulated by taking into consideration of interference constraints while guaranteeing the rate requirements of SUs and optimizing the time slot structure. Thus, the general optimization problem ($\textbf{P1}$) can be given as
\begin{equation} \label{GeneralOptimization}
\begin{array}{l}
\begin{array}{*{20}{c}}
{\mathop {\max }\limits_{{\theta_i},{f_{i,j}}} }&{\sum\limits_{i = 1}^K {\sum\limits_{j = 1}^M {{f_{i,j}}(1 - {\theta_i} - \frac{{{\tau _{{s_i}}}}}{T})lo{g_2}(1 + {H_{i,j}}\frac{{{\chi_i}{\theta_i}T - {\epsilon_{{s_i}}}}}{{T - {\theta_i}T - {\tau _{{s_i}}}}})} } }
\end{array}\\
\begin{array}{*{20}{l}}
{s.t.}&{{\bf{C1}}}&{{\chi_i}{\theta_i}T - {\epsilon_{{s_i}}} > 0,~\forall i \in {\cal K},}\\
{}&{{\bf{C2}}}&{T - {\theta_i}T - {\tau _{{s_i}}} > 0,~\forall i \in {\cal K},}\\
{}&{{\bf{C3}}}&{\sum\limits_{i \in {\mathcal{K}}} {\sum\limits_{j \in \mathcal{M}} {{f_{i,j}}\frac{{{\chi_i}{\theta_i}T - {\epsilon_{{s_i}}}}}{{T - {\theta_i}T - {\tau _{{s_i}}}}}{I_{i,j,m}} \le {I_{m}^{th}},} } }~\forall m \in {\cal L}\\
{}&{{\bf{C4}}}&{\sum\limits_{i \in {\mathcal{K}_\mathcal{R}}} {\sum\limits_{j \in \mathcal{M}} {{f_{i,j}}{r_{i,j}} \geqslant R_i^{req},} } } \\
{}&{{\bf{C5}}}&{\sum\limits_{i \in {\mathcal{K}_\mathcal{N}}} {\sum\limits_{j \in \mathcal{M}} {{f_{i,j}}{r_{i,j}} \geqslant {\zeta _i},} } } \\
{}&{{\bf{C6}}}&{\sum\limits_{i \in {\mathcal{K}}} {{f_{i,j}} = 1,~\forall j \in {\cal M},} }\\
{}&{{\bf{C7}}}&{{f_{i,j}} \in \{ 0,1\} ,~\forall i \in {\cal K},~\forall j \in {\cal M},}\\
{}&{{\bf{C8}}}&{0 < {\theta_i} < 1,~\forall i \in {\cal K},}
\end{array}
\end{array}
\end{equation}
where \textbf{C1} imposes energy of sensing to be less than the total harvested energy. \textbf{C2} means that the remaining time for data transmission must be greater than the sum of harvesting time and sensing time in one time slot. \textbf{C3} specifies that the total interference to the $m^{th}$ PU must be less than a given threshold. \textbf{C4} implies that the minimum required rate of RT SUs must be satisfied. \textbf{C5} means the NRT SUs rate must be greater than a given rate constraint. \textbf{C6} and \textbf{C7} specify that each sub-channel cannot be allocated to more than one SU. \textbf{C8} means the harvesting ratio should be a fraction of a time slot.

\section {Throughput Optimization} \label{Sub-Alloc}

\subsection{Solution Methodology}
Note that \textbf{P1} is an MINLP problem, which contains both binary variables $f_{i,j}$ and continuous variables $\theta_i$  for optimization. In fact, the objective function of the general optimization problem is not jointly convex for $\left\{ {{\theta _i},{f_{i,j}}} \right\}$. The MINLP optimization problems are  generally difficult to solve due to the combinatorial nature of mixed-integer programming (MIP) and the difficulty in solving nonlinear programming (NLP) problems \cite{00223}. 

Some methods such as outer-approximation methods, branch-and-bound, extended cutting plane methods, and the \textit{sorting and removing} method \cite{00223,00183-14} have been proposed to solve MINLP problems. However, the aforementioned methods cannot be exploited to our problem specific structures and properties. The minimax convex relaxation technique [33] can also be considered as a possible solution for MINLP problems. However, it cannot be applied to solve Eq.~(\ref{GeneralOptimization}) because it is not efficient for a large number of decision variables.

\begin{remark}
	The joint channel allocation and structure optimization problem (\textbf{P1}) is an MINLP problem, which exhibits the combinatorial nature of MIP problems and the difficulty in solving NLP problems. In fact, both MIP and NLP are considered NP-complete, and thus the joint resource allocation and structure optimization problem is NP-hard and requires exponential time complexity to achieve optimal solutions \cite{00223,00227,00238}.
\end{remark}

Since the general optimization problem is computationally intractable, a two-stage approach is considered for reducing complexity of the problem. This technique has achieved success in various scenarios \cite{00109}. Specifically, we first propose a sub-channel allocation scheme based on a factor called Energy Figure of Merit ($\alpha_{EFM}$). In this sub-channel allocation scheme, the heterogeneous SUs' rate requirements are roughly satisfied. Then, after removing the integer constraints of Eq. (\ref{GeneralOptimization}), the general nonconvex problem can be reduced to a new convex optimization problem. Thus, the optimum fraction of the time slot that each SU can harvest energy from the environment, can be obtained by solving the new convex optimization problem.

\subsection{Sub-channel Allocation Scheme} 

We focus on solving the general optimization problem \textbf{P1} by first employing the primal decomposition method where it can be decomposed into two subproblems. By having fixed the harvesting ratios ($\theta_i$) of each time slot, \textbf{P1} is simplified to the following optimization subproblem 

\begin{equation} \label{SubchannelFormula}
\begin{gathered}
\begin{array}{*{20}{c}}
\textbf{P2 :}~{\mathop {\max }\limits_{{f_{i,j}}} }&{\sum\limits_{i \in \mathcal{K}} {\sum\limits_{j \in \mathcal{M}} {{f_{i,j}}{r_{i,j}}} } } 
\end{array} \hfill \\
\begin{array}{*{20}{l}}
{s.t.}&{{\mathbf{C1}}}&{\sum\limits_{i \in \mathcal{K}} {\sum\limits_{j \in \mathcal{M}} {{f_{i,j}}{p_{i,j}}{I_{i,j,m}} \leqslant I_m^{th},~\forall m \in \mathcal{L},} } } \\ 
{}&{{\mathbf{C2}}}&{\sum\limits_{i \in {\mathcal{K}_\mathcal{R}}} {\sum\limits_{j \in \mathcal{M}} {{f_{i,j}}{r_{i,j}} \geqslant R_i^{req},} } } \\ 
{}&{{\mathbf{C3}}}&{\sum\limits_{i \in {\mathcal{K}_\mathcal{N}}} {\sum\limits_{j \in \mathcal{M}} {{f_{i,j}}{r_{i,j}} \geqslant {\xi _i},} } } \\ 
{}&{{\mathbf{C4}}}&{\sum\limits_{i \in \mathcal{K}} {{f_{i,j}}}  = 1,~\forall j \in {\cal M},} \\ 
{}&{{\mathbf{C5}}}&{{f_{i,j}} \in \{ 0,1\} ,~\forall i \in {\cal K},~\forall j \in {\cal M},} 
\end{array} \hfill \\ 
\end{gathered}
\end{equation}
where $p_{i,j}$ denotes the transmission power allocated by the $i^{th}$ SU to the $j^{th}$ available sub-channel.\\ 
Denote the total transmission power for each SU $\forall i \in {\cal K}$ as $\frac{{{\chi_i}{\theta_i}T - {\epsilon_{{s_i}}}}}{{T - {\theta_i}T - {\tau _{{s_i}}}}}$. It can be observed from the derivative of the transmission power with respect to $\theta_i$ that the transmission power is strictly increasing in $\theta_i \in (0,1)$, $\forall i \in {\cal K}$. Hence, the maximum transmission power occurs at the upper bound of the harvesting ratio ($\theta_i \simeq 1$). Meanwhile, initial harvesting ratios (initial transmission power) would not likely yield the maximum transmission power, and thus the interference limit ${\sum\limits_{i \in \mathcal{K}} {\sum\limits_{j \in \mathcal{M}} {{f_{i,j}}{p_{i,j}}{I_{i,j,m}} \leqslant I_m^{th}}}}$ is satisfied and can be ignored in the sub-channel allocation process. To solve \textbf{P2}, one of the most important considerations in the channel allocation process is to scrutinize the rate requirements of both RT and NRT cognitive users. Hence, the rate requirements in $\mathbf{C2}$ and $\mathbf{C3}$ play a key role in the sub-channel allocation process.

The following sub-channel allocation algorithm (Algorithm 1) requires $\theta_i$ to be initialized. Since ${\theta _i} > \frac{{{\epsilon_{{s_i}}}}}{{{\chi _i}T}}$ and ${\theta _i} < \frac{{T - {\tau _{{s_i}}}}}{T}$, $\forall i \in \mathcal{K}$, the initial $\theta_i$, $\forall i \in \mathcal{K}$ can be expressed as

\begin{equation} \label{initital}
\theta _i^{initial} = \frac{{{\epsilon _{{s_i}}}}}{{2{\chi _i}T}} + \frac{1}{2}\left( {1 - \frac{{{\tau _{{s_i}}}}}{T}} \right),
\end{equation}
i.e., the average of the upper and lower bounds.

\begin{algorithm}[H] \label{algorithm}
	\textbf{Initialization:}\\
	Initial rates of SUs: ${\cal R} = \{ {R_1},...,{R_i}\}  = 0$\\
	EFM factor for each SU:	$\alpha _{EFM}^i = \frac{{{\chi_i}}}{{{\epsilon_{{s_i}}}}},\forall i \in {\cal K}$\\
	${{\cal M}_t} = {\cal M},{{\cal D}_i} = \emptyset ,\forall i \in {\cal K}$\\
	$\theta _i^{(0)}, \forall i \in \mathcal{K},$ have the initial values in Eq.~(\ref{initital})\\
	\textbf{Sub-channel Allocation for RT SUs:}\\
	{While ${{\cal M}_t} \ne \emptyset $ and $\min ({R_i} - R_i^{req}) < 0$ }\\{
		Find ${i^*}$ that satisfies $\alpha _{EFM}^{{i^*}} \ge \alpha _{EFM}^i$ for $\forall i \in {\cal K_{R}}$;\\
		Finding the best sub-channel for the ${i^*}$:\\
		Find ${j^*}:$ ${j^*} = \arg \mathop {\max }\limits_{j \in {{\cal M}_t}} r_{{i^*},{j^*}}$;\\
		${{\cal M}_t} = {{\cal M}_t}/{j^*}$ and ${{\cal D}_{{i^*}}} = {{\cal D}_{{i^*}}} \cup {j^*}$;\\
		${R_{{i^*}}} = {R_{{i^*}}} + { (1 - {\theta_{i^*}^{(0)}} - \frac{{{\tau _{{s_{i^*}}}}}}{T})lo{g_2}(1 + {H_{{i^*},{j^*}}}\frac{{{\chi_{i^*}}{\theta_{i^*}^{(0)}}T - {\epsilon_{{s_{i^*}}}}}}{{T - {\theta_{i^*}^{(0)}}T - {\tau _{{s_{i^*}}}}}})}$;\\
		End while
	}\\
	\textbf{Define ${\cal K}_R^{Al} = \{ i \in {{\cal K}_R},{{\cal D}_{{i}}} \ne \emptyset \}$}\\
	\textbf{Sub-channel Allocation for NRT SUs:}\\
	{While ${{\cal M}_t} \ne \emptyset $ }\\{
		Find ${i^*}$ that satisfies $\alpha _{EFM}^{{i^*}} \ge \alpha _{EFM}^i$ for $\forall i \in {\cal K_{N}}$;\\
		Finding the best sub-channel for the ${i^*}$:\\
		Find ${j^*}:$ ${j^*} = \arg \mathop {\max }\limits_{j \in {{\cal M}_t}} r_{{i^*},{j^*}}$;\\
		${{\cal M}_t} = {{\cal M}_t}/{j^*}$ and ${{\cal D}_{{i^*}}} = {{\cal D}_{{i^*}}} \cup {j^*}$;\\
		${R_{{i^*}}} = {R_{{i^*}}} + r_{{i^*},{j^*}}$;\\
		End while
	}\\
	\textbf{Define ${\cal K}_N^{Al} = \{ i \in {{\cal K}_N},{{\cal D}_{{i}}} \ne \emptyset \}$}\\
	\caption{\textbf{E}nergy \textbf{F}igure of \textbf{M}erit (EFM) based sub-channel allocation}
\end{algorithm}

In the sub-channel allocation process, RT SUs have higher priority for sub-channel allocations as compared to NRT SUs. Thus, the sub-channel allocation would be done for RT SUs until the minimum rate requirements of RT SUs are satisfied. During each cycle, RT SU whose EFM factor ($\alpha _{EFM}^i = \frac{{{\chi_i}}}{{{\epsilon_{{s_i}}}}}$) is greater than the others has the priority to get a sub-channel among the available ones. In fact, the higher $\alpha_{EFM}$ stems from a greater amount of energy extracted from the environment and less amount of energy consumed by the spectrum sensing process. Thus, those SUs having higher $\alpha_{EFM}$ normally need less numbers of sub-channels to meet the required rate. Furthermore, the chosen SU preferably receives a sub-channel that has the highest corresponding achievable rate.

After RT SUs have been assigned sub-channels, the remaining ones are allocated to NRT SUs to meet their rate constraints. Like the RT sub-channel allocation process, the sub-channel assignments for NRT SUs follow the EFM-based user preference. The sub-channel allocation scheme continues until all sub-channels are assigned to SUs. Note that at the end of each round of sub-channel allocation for RT and NRT SUs, we define a new set for those SUs which have been assigned sub-channels. ${\cal K}_R^{Al}$ and ${\cal K}_N^{Al}$ are sets of RT and NRT SUs that have received sub-channels, respectively, and ${{\cal K}^{Al}} = {\cal K}_R^{Al} \cup {\cal K}_N^{Al}$ is the set of all SUs with allocated sub-channels. In this work, each SU is assumed to transmit all its harvested power over all its allocated sub-channels. The power allocation procedure for the assigned sub-channels to SUs is beyond the scope of this paper.

\subsection{Structure Optimization}

After sub-channel allocation, the integer constraints of Eq. (\ref{GeneralOptimization}) are removed because binary variables $f_{i,j}$ take on 0 or 1 indicating whether sub-channels are allocated or not. Thus, the new optimization problem, which aims at maximizing the sum-rate of SUs by finding optimum fractions of harvesting for all SUs, can be expressed as 
\begin{equation} \label{ConvexOpt}
\begin{split}
\begin{array}{l}
\begin{array}{*{20}{c}}
\textbf{P3:}{\mathop {\max }\limits_{{\theta_i}} }~{\sum\limits_{i \in {{\cal K}^{Al}}} {\sum\limits_{j \in {{\cal D}_i}} {(1 - {\theta_i} - \frac{{{\tau _{{s_i}}}}}{T})lo{g_2}(1 + {H_{i,j}}\frac{{{\chi_i}{\theta_i}T - {\epsilon_{{s_i}}}}}{{T - {\theta_i}T - {\tau _{{s_i}}}}})} } }
\end{array}\\
\begin{array}{*{20}{l}}
{s.t.}&{{\bf{C1}}}&{{\chi_i}{\theta_i}T - {\epsilon_{{s_i}}} > 0,~\forall i \in {\cal K}^{Al},}\\
{}&{{\bf{C2}}}&{T - {\theta_i}T - {\tau _{{s_i}}} > 0,~\forall i \in {\cal K}^{Al},}\\
{}&{{\bf{C3}}}&{\sum\limits_{i \in{\cal K}^{Al}} {\sum\limits_{j \in {{\cal D}_i}} {\frac{{{X_i}{\theta_i}T - {\epsilon_{{s_i}}}}}{{T - {\theta_i}T - {\tau _{{s_i}}}}}{I_{i,j,m}} \le {{I_{m}^{th}}},} }~\forall m \in {\cal L}, }\\
{}&{{\bf{C4}}}&{\sum\limits_{j \in {{\cal D}_i}} {r_{i,j}} \ge R_i^{req},~\forall i \in {\cal K}_R^{Al}, }\\
{}&{{\bf{C5}}}&{\sum\limits_{j \in {{\cal D}_i}} {r_{i,j}} \ge {\gamma _i} ,~\forall i \in {\cal K}_N^{Al}, }\\
{}&{{\bf{C6}}}&{0 < {\theta_i} < 1,~\forall i \in {\cal K}^{Al}.}
\end{array}
\end{array}
\end{split}
\end{equation}
If \textbf{P3} describes a convex optimization problem, it can be solved by standard convex optimization methods such as the barrier method or iterative gradient technique with duality. Hence, the optimum fractions of the time slot for energy harvesting of SUs can be obtained. Therefore, since it is important to analyze convexity of \textbf{P3}, convexity of the objective function is established by the following Lemma.

\begin{lemma} \label{[Concavity]}
	The objective function in \textbf{P3} for $\forall i \in \mathcal{K}$, $0 < {\theta _i} < 1,$  is a concave function.
\end{lemma}
\begin{IEEEproof}
	The Lemma is proved in Appendix A.
\end{IEEEproof}
Having proven the convexity of \textbf{P3}, the optimal harvesting ratios can be obtained; however, the constraints should be examined. In general, if constraint functions in an MINLP optimization problem are convex, the optimization problem is called a convex MINLP \cite{00233}.
\begin{lemma} \label{[CVXMINLP]}
	The joint resource allocation and time slot optimization problem \textbf{P1} is a convex MINLP. 
\end{lemma}
\begin{IEEEproof}
	The Lemma is proved in Appendix B.
\end{IEEEproof}
Having proven \textbf{P1} being convex MINLP, one can propose a heuristic algorithm (like Feasibility Pump\footnote{The Feasibility Pump (FP) is one of the most well known primal heuristic for mixed integer non-linear programming \cite{00233}.} (FP)) to obtain the optimal solution. The FP algorithm decomposes a mathematical programming problem into two parts: integer feasibility and constraint feasibility. For the convex MINLP scenario, the solution can be achieved by solving an LP or a convex NLP, which can be done in polynomial time \cite{00233}.

To obtain the optimal solution of \textbf{P3}, the associated Lagrangian can be expressed as 
\begin{equation} \label{lagrangian}
\begin{gathered}
L({\theta _1},{\theta _2},...,{\theta _K},{\lambda _i},{\mu _i},{\nu _m},\rho _i^R,\rho _i^N) =  \hfill \\
- \sum\limits_{i \in {\mathcal{K}^{Al}}} {\sum\limits_{j \in {\mathcal{D}_i}} {\left( {1 - {\theta _i} - \frac{{{\tau _{{s_i}}}}}{T}} \right)lo{g_2}\left( {1 + {H_{i,j}}\frac{{{\chi _i}{\theta _i}T - {\varepsilon _{{s_i}}}}}{{T - {\theta _i}T - {\tau _{{s_i}}}}}} \right)} }  \hfill \\
+ {\lambda _i}({\varepsilon _{{s_i}}} - {\chi _i}{\theta _i}T) + {\mu _i}({\theta _i}T + {\tau _{{s_i}}} - T) \hfill \\
+ {\nu _m}\left( {\sum\limits_{i \in {\mathcal{K}^{Al}}} {\frac{{{\chi _i}{\theta _i}T - {\varepsilon _{{s_i}}}}}{{T - {\theta _i}T - {\tau _{{s_i}}}}}{I_{i,m}} - I_m^{th}} } \right) \hfill \\
+ \rho _i^R\left( \begin{gathered}
R_i^{req} -  \hfill \\
\sum\limits_{j \in {\mathcal{D}_i}} {\left( {1 - {\theta _i} - \frac{{{\tau _{{s_i}}}}}{T}} \right)} lo{g_2}\left( {1 + {H_{i,j}}\frac{{{\chi _i}{\theta _i}T - {\varepsilon _{{s_i}}}}}{{T - {\theta _i}T - {\tau _{{s_i}}}}}} \right) \hfill \\ 
\end{gathered}  \right)\\
+ \rho _i^N\left( \begin{gathered}
{\gamma _i} -  \hfill \\
\sum\limits_{j \in {\mathcal{D}_i}} {\left( {1 - {\theta _i} - \frac{{{\tau _{{s_i}}}}}{T}} \right)} lo{g_2}\left( {1 + {H_{i,j}}\frac{{{\chi _i}{\theta _i}T - {\varepsilon _{{s_i}}}}}{{T - {\theta _i}T - {\tau _{{s_i}}}}}} \right) \hfill \\ 
\end{gathered}  \right)
\end{gathered}
\end{equation}
where ${\lambda _i}$, ${\mu _i}$, ${\nu _m}$, ${\rho _i^R}$ and ${\rho _i^N}$ are Lagrange multipliers. The dual function of \textbf{P3} in Eq. (\ref{ConvexOpt}) is
\begin{equation} \label{DualFunction}
g(\lambda ,\mu ,\nu ,\rho^R ,\rho^N ) = \mathop {sup}\limits_{{\theta _1},...,{\theta _k}} L({\theta _1},...,{\theta _k},\lambda ,\mu ,\nu ,\rho^R ,\rho^N).
\end{equation}
Thus, the dual optimization problem for Eq. (\ref{ConvexOpt}) is
\begin{equation} \label{DualOptimization}
\begin{array}{*{20}{c}}
\textbf{P4 :}~{\min }&{g(\lambda ,\mu ,\nu ,\rho^R ,\rho^N)} \\ 
{}&{\lambda  \succ 0,\mu  \succ 0,\nu  \succ 0,\rho^R  \succ 0,\rho^N  \succ 0} 
\end{array}
\end{equation}
In order to solve the dual problem, an iterative scheme using the gradient projection method can be applied. Thus, the Lagrange multiplier for \textbf{C1} is given as
\begin{equation}
{\lambda_i ^{(t + 1)}} = {\left[ {{\lambda_i ^{(t)}} - {\alpha ^{(t)}}\frac{{dL}}{{d\lambda_i }}} \right]^ + } = {\left[ {{\lambda_i ^{(t)}} - {\alpha ^{(t)}}({\epsilon_{{s_i}}} - \chi_i \theta_i T)} \right]^ + }. \hfill
\end{equation}
The Lagrange multiplier for \textbf{C2} can be written as
\begin{equation}
{\mu_i ^{(t + 1)}} = {\left[ {{\mu_i ^{(t)}} - {\beta ^{(t)}}\frac{{dL}}{{d\mu_i }}} \right]^ + } = {\left[ {{\mu_i ^{(t)}} - {\beta ^{(t)}}(\theta_i T + {\tau_{{s_i}}} - T)} \right]^ + }. \hfill
\end{equation}
Likewise, the Lagrange multiplier of \textbf{C3} is expressed as
\begin{equation} 
\begin{gathered}
\nu _m^{(t + 1)} = {\left[ {\nu _m^{(t)} - {\pi ^{(t)}}\frac{{dL}}{{d\nu_m }}} \right]^ + } =  \hfill \\
{\left[ {\nu _m^{(t)} - {\pi ^{(t)}}\left( {\sum\limits_{i \in {\mathcal{K}^{Al}}} {\sum\limits_{j \in {\mathcal{D}_i}} {\frac{{{\chi _i}{\theta _i}T - {\epsilon _{{s_i}}}}}{{T - {\theta _i}T - {\tau _{{s_i}}}}}{I_{i,j,m}} - I_m^{th}} } } \right)} \right]^ + }. \hfill \\ 
\end{gathered}
\end{equation}
The Lagrange multipliers of transmission rates for \textbf{C4} and \textbf{C5} are
\begin{equation}
\begin{split}
{\rho_i ^{R,(t + 1)}} =& {\left[ {{\rho_i ^{R,(t)}} - {\psi ^{(t)}}\frac{{dL}}{{d\rho_i ^{R} }}} \right]^ + } =  \hfill \\
&{\left[ {{\rho_i ^{R,(t)}} - {\psi ^{(t)}}\left( {R_i^{req} - \sum\limits_{j \in {\mathcal{D}_i}} {{r_{i,j} (\theta_i)}} } \right)} \right]^ + }, \hfill \\ 
\end{split}
\end{equation}
and
\begin{equation}
\begin{split}
{\rho_i ^{N,(t + 1)}} =& {\left[ {{\rho_i ^{N,(t)}} - {\eta ^{(t)}}\frac{{dL}}{{d\rho_i ^{N} }}} \right]^ + } =  \hfill \\
&{\left[ {{\rho_i ^{N,(t)}} - {\eta ^{(t)}}\left( {{\gamma _i} - \sum\limits_{j \in {\mathcal{D}_i}} {{r_{i,j}(\theta_i)}} } \right)} \right]^ + }, \hfill \\ 
\end{split}
\end{equation}
where $t$ is the iteration index, ${\alpha ^{(t)}}$, ${\beta ^{(t)}}$, ${\pi ^{(t)}}$, ${\psi ^{(t)}}$ and ${\eta ^{(t)}}$ are sufficiently small positive step-sizes, and ${[a]^ + } = \max (0,a)$. 
\begin{proposition} \label{Theo1}
	Dual variables ${\lambda_i ^{(t)}}$, ${\mu_i ^{(t)}}$, ${\nu_m ^{(t)}}$, ${\rho_i ^{R,(t)}}$, and ${\rho_i ^{N,(t)}}$ can eventually converge to the dual optimal solution ${\lambda}$, ${\mu}$, ${\nu}$, ${\rho}$, and ${\varphi}$ if the step sizes are chosen such that ${\alpha ^t} \to 0,\sum\limits_{t = 0}^\infty  {{\alpha ^t} = \infty } ,$ ${\beta ^t} \to 0,\sum\limits_{t = 0}^\infty  {{\beta ^t} = \infty } ,$ ${\pi ^t} \to 0,\sum\limits_{t = 0}^\infty  {{\pi ^t} = \infty } ,$ ${\psi ^t} \to 0,\sum\limits_{t = 0}^\infty  {{\psi ^t} = \infty } ,$ and ${\eta ^t} \to 0,\sum\limits_{t = 0}^\infty  {{\eta ^t} = \infty } .$
\end{proposition}
\begin{IEEEproof}
	The proof is presented in Appendix C.
\end{IEEEproof}

The primal problem Eq. (\ref{ConvexOpt}) is convex with positive constraints (as verified in Lemma 1 and 2). Thus, one can conclude from Proposition 1 that the Slater's condition for strong duality of the primal problem holds (duality gap is zero). Thus, the optimal solution to Eq. (\ref{DualOptimization}) is the global maximum of the primal problem. To obtain the optimal solution of harvesting ratios for SUs, the following subgradient method is used.

\begin{algorithm}[H] \label{algorithm2}
	\textbf{Initialization:}\\
	Setting $\theta _i$\\
	Initialize $\lambda _i^{(0)}$, $\mu _i^{(0)}$, $\rho _i^{R,(0)}$, $\rho _i^{N,(0)}$, and $\nu _m^{(0)} = 0$\\
	\textbf{Repeat} for $t \ge 1$\\
	Compute optimal harvesting ratios:\\
	$\theta _i^{*} = \arg \mathop {\max }\limits_{{\theta _i}} (\lambda _i^{(t)},\mu _i^{(t)},\rho _i^{R,(t)},\rho _i^{N,(t)},\nu_m^{(t)})$\\
	Update dual variables:\\
	${\lambda_i ^{(t + 1)}} = {[{\lambda_i ^{(t)}} - {\alpha ^{(t)}}\frac{{dL}}{{d\lambda_i }}]^ + }$, ${\mu_i ^{(t + 1)}} = {[{\mu_i ^{(t)}} - {\beta ^{(t)}}\frac{{dL}}{{d\mu_i }}]^ + }$, ${\rho_i ^{R,(t + 1)}} = {[{\rho_i ^{R,(t)}} - {\psi ^{(t)}}\frac{{dL}}{{d\rho_i^R }}]^ + }$, ${\rho_i ^{N,(t + 1)}} = {[{\rho_i ^{N,(t)}} - {\eta ^{(t)}}\frac{{dL}}{{d\rho_i^N }}]^ + } $, and ${\nu_m ^{(t + 1)}} = {[{\nu_m ^{(t)}} - {\pi ^{(t)}}\frac{{dL}}{{d\nu_m }}]^ + }$\\
	\textbf{Until} Convergence\\
	\caption{Structure Optimization By Iterative Gradient Method}
\end{algorithm}
Apart from solving the convex optimization problem using iterative gradient method, the optimal solution can be obtained by deriving a closed-form solution. The following theorem proposes a closed-form solution for optimal harvesting ratios of SUs solved by the AP.

\begin{theorem} \label{LambertTheorem}
	The optimal solution of the harvesting ratio in one time slot for each SU, $\forall i \in \mathcal{K} = \{ 1,2,...,K\},$ can be attained as:\\
	\begin{equation} \label{TheoremGlobalOptimal}
	\theta _i^* = \frac{{T - {\tau _{{s_i}}}}}{T} - \frac{{\mathcal{W}\left( {\frac{{H{\chi _i} - 1}}{e}} \right)H\left( {{\chi _i}T - {\chi _i}{\tau _{{s_i}}} - {\varepsilon _{{s_i}}}} \right)}}{{T\left( {H{\chi _i} - 1} \right)\left( {1 + \mathcal{W}\left( {\frac{{H{\chi _i} - 1}}{e}} \right)} \right)}},
	\end{equation}
	where $\mathcal{W} (.)$ refers to the Lambert $\mathcal{W}$ function \cite{147-17}.
\end{theorem}

\begin{IEEEproof}
	We begin with the assumption that the harvesting ratio allocation set of Eq.~(\ref{ConvexOpt}) is a nonempty, convex and compact set \cite{AbbasToward242}. Hence, the objective function is strictly concave with respect to $\theta_{i}$. 
	Let ${\lambda _i},{\mu _i} \geqslant 0$, $\forall i \in \mathcal{K} = \{ 1,2,...,K\}$, ${\nu _m} \geqslant 0$, $\forall m \in \mathcal{L} = \{ 1,2,...,L\}$, $\rho _i^R \geqslant 0$, $\forall i \in {\mathcal{K}_R} = \{ {i_0} + 1,...,K\}$, and $\rho _i^N \geqslant 0$, $\forall i \in {\mathcal{K}_N} = \{ 1,2,...,{i_0}\}$ denote the Lagrange multipliers of lower-bound energy and time of transmission in \textbf{C1} and \textbf{C2}, interference constraint in \textbf{C3}, lower-bound transmission rate for RT and NRT users in \textbf{C4} and \textbf{C5}, respectively.
	
	Therefore, using the Lagrangian of the optimization problem \textbf{P2} in Eq.~(\ref{lagrangian}), the objective function can be optimized by exploiting the necessary and sufficient conditions,
	\begin{equation}
	\nabla \mathcal{L}({\theta ^*},\lambda ,\mu ,\nu ,{\rho ^R},{\rho ^N}) = 0.
	\end{equation}
	The objective function is given as
	\begin{equation}
	Obj = \left( {\frac{{T - T{\theta _i} - {\tau _{{s_i}}}}}{{T\ln 2}}} \right)lo{g_2}\left( {1 + {H_{i,j}}\frac{{{\chi _i}{\theta _i}T - {\varepsilon _{{s_i}}}}}{{T - {\theta _i}T - {\tau _{{s_i}}}}}} \right) \hfill \\
	\end{equation}
	The derivative of the objective function with respect to the vector of harvesting ratios can be expressed as
	\begin{equation}
	\begin{split}
	\frac{d}{{d{\theta _i}}}\left( {Obj} \right) &= \frac{1}{{\ln 2}}\left[ {\ln \left( {1 + H\frac{{\chi \theta T - {\varepsilon _s}}}{{T - \theta T - {\tau _s}}}} \right) - } \right. \hfill \\
	&\left. {\frac{{H\chi \left( {T - \theta T - {\tau _s}} \right) + H\left( {\chi \theta T - {\varepsilon _s}} \right)}}{{\left( {T - \theta T - {\tau _s}} \right) + H\left( {\chi \theta T - {\varepsilon _s}} \right)}}} \right]. \hfill \\ 
	\end{split}
	\end{equation}
	Then, considering $\frac{{d\left( {\nabla \mathcal{L}} \right)}}{{d{\theta _i}}} = 0$, we have
	\begin{equation}
	\begin{gathered}
	\frac{1}{{\ln 2}}\left[ \begin{gathered}
	\ln \left( {1 + H\frac{{\chi \theta T - {\varepsilon _s}}}{{T - \theta T - {\tau _s}}}} \right) \hfill \\
	- \frac{{H\chi \left( {T - \theta T - {\tau _s}} \right) + H\left( {\chi \theta T - {\varepsilon _s}} \right)}}{{\left( {T - \theta T - {\tau _s}} \right) + H\left( {\chi \theta T - {\varepsilon _s}} \right)}} \hfill \\ 
	\end{gathered}  \right] - {\lambda _i}{\chi _i}T \hfill \\
	+ {\mu _i}T + {\nu _m}{I_m}\frac{{H\chi T\left( {T - \theta T - {\tau _s}} \right) + HT\left( {\chi \theta T - {\varepsilon _s}} \right)}}{{{{\left( {T - \theta T - {\tau _s}} \right)}^2}}} \hfill \\
	+ \rho _i^R\left( {\frac{d}{{d{\theta _i}}}\left( {Obj} \right)} \right) + \rho _i^N\left( {\frac{d}{{d{\theta _i}}}\left( {Obj} \right)} \right) = 0, \hfill \\ 
	\end{gathered}
	\end{equation}
	and the Lagrange multipliers constraints are equal to zero as follows
	\begin{equation} \label{mulitpliersalltogether}
	\left\{ {\begin{array}{*{20}{l}}
		{{\lambda _i}({\varepsilon _{{s_i}}} - {\chi _i}{\theta _i}T) = 0,}&{i \in \mathcal{K}} \\ 
		{{\mu _i}({\theta _i}T + {\tau _{{s_i}}} - T) = 0,}&{i \in \mathcal{K}} \\ 
		{{\nu _m}\left( {\frac{{{\chi _i}{\theta _i}T - {\varepsilon _{{s_i}}}}}{{T - {\theta _i}T - {\tau _{{s_i}}}}}{I_{i,m}} - I_m^{th}} \right) = 0,}&{i \in \mathcal{K}}, ~{m \in \mathcal{L}}\\
		{\rho _i^N\left( {{\gamma _i} - R_i^{NT}} \right) = 0,}&{i \in {\mathcal{K}_N}} \\ 
		{\rho _i^R\left( {R_i^{req} - R_i^{RT}} \right) = 0,}&{i \in {\mathcal{K}_R}}
		\end{array}} \right.
	\end{equation}
	\begin{figure}[t]
		\centering
		\includegraphics[width=18cm,height=8cm]{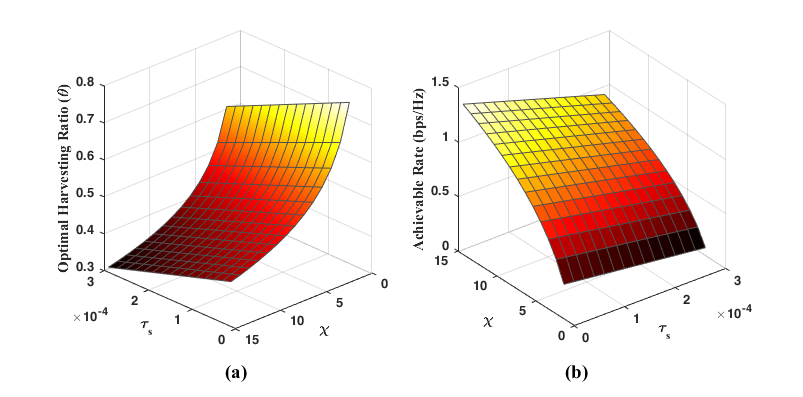}
		\caption{The performance evolution of the heterogeneous CR system for one SU scenario: a) optimal harvesting ratio versus energy harvesting rate and time of sensing; b) achievable rate versus energy harvesting rate and sensing time.} 
		\label{fig:3DFigAandB}
	\end{figure}
	In order to obtain the solutions, the values of multipliers in Eq.~(\ref{mulitpliersalltogether}) should be considered. The first constraint ${\varepsilon _{{s_i}}} = {\chi _i}{\theta _i}T$, represents the special case where the total harvested energy is consumed for the spectrum sensing. The second constraint ${\theta _i}T = T- {\tau _{{s_i}}}$ denotes the special case where there is no remaining time for data transmission. Thus, we are not interested in special cases and one can conclude ${\lambda _i}=0$ and ${\mu _i}=0$. The rate constraints, $R_i^{RT} = R_i^{req}$ and ${R_i^{NT} = {\gamma _i}}$, represent the special cases where the transmission rates of RT and NRT users meet the lower bound values which are not generally desired. For simplicity in the last constraint, we assume that the total interference introduced by SUs is always less than the maximum interference threshold. Therefore, one can conclude that ${\nu _m}=0$ because the total interference is assumed to be less than the threshold ($I_m^{th}$). Hence, the optimal harvesting time can be obtained by solving the following equation.
	\begin{equation} \label{ObjMosaviSefr}
	\begin{split}
	\frac{1}{{\ln 2}}&\left[ {\ln \left( {1 + H\frac{{\chi \theta T - {\varepsilon _s}}}{{T - \theta T - {\tau _s}}}} \right) } \right. \hfill \\
	&-\left. {\frac{{H\chi \left( {T - \theta T - {\tau _s}} \right) + H\left( {\chi \theta T - {\varepsilon _s}} \right)}}{{\left( {T - \theta T - {\tau _s}} \right) + H\left( {\chi \theta T - {\varepsilon _s}} \right)}}} \right]=0.\hfill \\ 
	\end{split}
	\end{equation}
	Define intermediate variables $a$ and $b$ as $a = T - \theta T - {\tau _s}$ and $b = \chi \theta T - {\varepsilon _s}$. Thus, we can re-write Eq.~(\ref{ObjMosaviSefr}) as
	\begin{equation} 
	\ln \left( {\frac{{a + Hb}}{a}} \right) = \frac{{H\chi a + Hb}}{{a + Hb}}.
	\end{equation}
	One can conclude that $b =  - \chi a + c,$ where $c$ is defined as $c = \chi T - \chi {\tau _s} - {\varepsilon _s}$. Therefore, 
	\begin{equation}
	\ln \left( {1 - H\chi  + \frac{{Hc}}{a}} \right) = \frac{{Hc}}{{a\left( {1 - H\chi } \right) + Hc}}.
	\end{equation}
	Define a new variable $t$ as $t = 1 - H\chi  + \frac{{Hc}}{a}$. Then,
	\begin{equation}\label{lntforlemma}
	t\ln \left( t \right) = t + H\chi  - 1.
	\end{equation}
	
	\begin{lemma} \label{lemma1}
		The solution of $x\ln \left( x \right) = ax + b$ ($a$ and $b$ are constants), is\\
		\[x = \frac{b}{{\mathcal{W}\left( {\frac{b}{e^{a}}} \right)}},\]
		where $\mathcal{W} (.)$ is the Lambert $\mathcal{W}$ function.
	\end{lemma}
	
	\begin{IEEEproof}
		The Lemma is proved in Appendix D.	
	\end{IEEEproof}
	Based on Lemma~\ref{lemma1}, the solution of Eq.~(\ref{lntforlemma}) can be expressed as
	\begin{equation}
	t = \frac{{H\chi  - 1}}{{\mathcal{W}\left( {\frac{{H\chi  - 1}}{e}} \right)}}.
	\end{equation}
	Therefore, using $t = 1 - H\chi  + \frac{{Hc}}{a}$ and $a = T - \theta T - {\tau _s}$, the global optimal solution Eq.~(\ref{TheoremGlobalOptimal}) is proved.\\
\end{IEEEproof}

\section{Simulation Results}\label{sec:simultion}

Simulations have been conducted to demonstrate the performance of our proposal. Meanwhile, the impact of various parameters like interference thresholds of PUs, transmission rate constraints, energy harvesting rate, and sensing time on the network performance have been analyzed. 

\subsection{Simulation Setup} \label{sec:SimulSetup}
In all the simulations, channel gains are modeled as $h_{i,j} = {\mathcal{Y}}d_{i,j}^{-\beta }$, where $\mathcal{Y}$ is a random value generated according to the Rayleigh distribution, $d_{i,j}^{-\beta}$ is the geographical distance between the transmitter and receiver, and $\beta$ is the path-loss exponent \cite{00136}. $d$ varies between 50~m to 200~m, and $\beta=3$. The bandwidth of each OFDM sub-channel is 62.5 kHz, and the noise power is $10^{-13}$W (or -100 dbm) in our simulation analysis. The overall probabilities of PUs' detection, mis-detection and false alarm are uniformly distributed over [0,1], [0.01, 0.05], and [0.05, 0.1], respectively. Various experimental results are provided to deeply analyze the performance of our network and investigate effects of different system parameters. 

\subsection{One SU Scenario}
In this part, an experiment has been conducted to evaluate the performance of our system versus different system parameters. Fig. \ref{fig:3DFigAandB}(a) illustrates the optimal harvesting ratio ($\theta$) versus various amounts of harvesting rate ($\chi$) and spectrum sensing time for one SU ($\epsilon_s=1$ mJ). It is clearly shown that a larger harvesting fraction is preferred when the spectrum sensing time decreases. At the same time as energy harvesting rate of the SU declines, the harvesting ratio grows exponentially. Fig. \ref{fig:3DFigAandB}(b) demonstrates the achievable throughput versus different sensing times and energy harvesting rates for one SU over a single sub-channel ($\epsilon_s=1$ mJ). It can be seen that the achievable throughput experiences a sharp increase as the energy harvesting rate improves. Meanwhile, the higher achievable throughput is accomplished by decreasing the sensing time because more time is left for data transmission.

\subsection{Sub-channel Allocation Performance}
\begin{figure}[t]
	\centering
	\includegraphics[width=9cm,height=7cm]{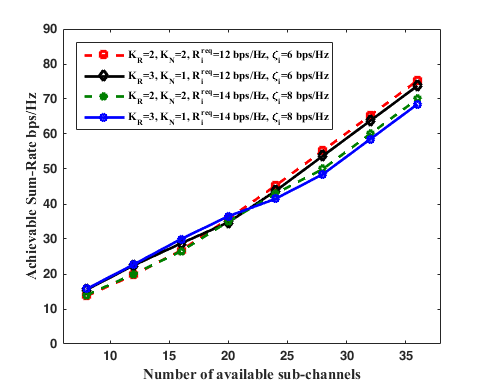}
	\caption{The achievable sum-rate as a function of the number of available sub-channels.}
	\label{fig:SumRateSubChannels}
\end{figure}

To evaluate the performance of our proposed sub-channel allocation algorithm, a series of experiments have been conducted. Fig. \ref{fig:SumRateSubChannels} evaluates the sum rate of a CR network ($K=4$ SUs) for different numbers of available sub-channels. Note that the achievable sum rate of the CR system increases as the number of available sub-channels grows. Consider four scenarios with 2 RT and 2 NRT users for two cases and 3 RT and 1 NRT users for the others. SU-1 and SU-2 are RT users whose energy harvesting rates are $\chi_1=30$ mJ/s and $\chi_2=40$ mJ/s, respectively. SU-3 and SU-4 have the energy harvesting rates of $\chi_3=60$ mJ/s and $\chi_4=120$ mJ/s, respectively. However, SU-4 is always an NRT user while SU-3 can be RT or NRT in different cases. As shown in Fig. \ref{fig:SumRateSubChannels}, the achievable sum rate not only is related to the number of sub-channels and number of RT and NRT users, but also depends on the required rates of RT SUs and rate constraints of NRT users. When $R^{req}$ is higher, RT users which have lower harvesting rates need more sub-channels to meet their required rates. More specifically, in Fig. \ref{fig:SumRateSubChannels}, as the required rate increases from $R^{req}=12$ bps/Hz to $R^{req}=14$ bps/Hz and the rate constraint grows from $\zeta=6$ bps/Hz to $\zeta=8$ bps/Hz, the achievable sum rate slightly decreases for higher number of available sub-channels.

\begin{figure}[t]
	\centering
	\includegraphics[width=9cm,height=7cm]{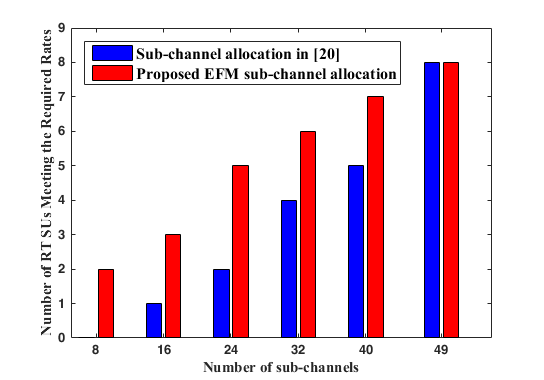}
	\caption{Number of RT SUs successfully meeting their required rates versus number of available OFDM sub-channels.}
	\label{fig:EFMComparison}
\end{figure}

As a second illustrative example, simulation results of a CR network with 8 RT SUs are illustrated in Table 3. In this part, we assume that the channel gains are identical for all SUs. It is shown that by increasing the value of $\alpha_{EFM}$ for different SUs, their single achievable rates improve. Thus, the RT SUs with higher $\alpha_{EFM}$ require less sub-channels to achieve the required rate. Fig. \ref{fig:EFMComparison} illustrates the comparison between our EFM-based sub-channel allocation in Section \ref{Sub-Alloc} and the sub-channel allocation scheme in~\cite{00109}. Fig. \ref{fig:EFMComparison} explicitly shows that the number of RT users who meet their required rates are significantly higher, especially for a small number of available sub-channels. In other words, since in the EFM-based method, the RT SUs with higher $\alpha_{EFM}$ have higher priority for sub-channel allocation, the SUs with less required sub-channels receive sub-channels first. Whereas, the method in~\cite{00109} does not consider this prioritized criterion regarding RT SUs sub-channel allocation.\\
\begin{table} \label{Example}
	\centering
	\caption{Simulation parameters for an experiment of 8 SUs.}
	\begin{tabular}{|c c c c c c c|} 
		\hline
		SU&$\alpha_{EFM}$&$\chi$&$\theta_{opt}$&$r$&$R^{req}$&Req. Sub-Chan. \\ [0.5ex] 
		\hline\hline
		1 & 3060 & 20 & 0.554 &1.00 & 10 bps/Hz & 10 sub-channels  \\ 
		\hline
		2 & 5850 & 20 & 0.455 &1.25 & 10 bps/Hz & 8 sub-channels \\
		\hline
		3 & 7230 & 30 & 0.412 &1.50 & 10 bps/Hz & 7 sub-channels\\
		\hline
		4 & 10130 & 40 & 0.363 &1.75 & 10 bps/Hz & 6 sub-channels\\
		\hline
		5 & 12500 & 60 & 0.336 &2.00 & 10 bps/Hz & 5 sub-channels\\ 
		\hline
		6 & 15560 & 85 & 0.309 &2.25 & 10 bps/Hz & 5 sub-channels\\ 
		\hline
		7 & 19050 & 120& 0.285 &2.50 & 10 bps/Hz & 4 sub-channels\\ 
		\hline
		8 & 31000 & 160& 0.258 &2.75 & 10 bps/Hz & 4 sub-channels\\ 
		\hline
	\end{tabular}
\end{table}

\subsection{Structure Optimization for Fixed Sub-channel Allocations}
\begin{figure}[t]
	\centering
	\includegraphics[width=9cm,height=7cm]{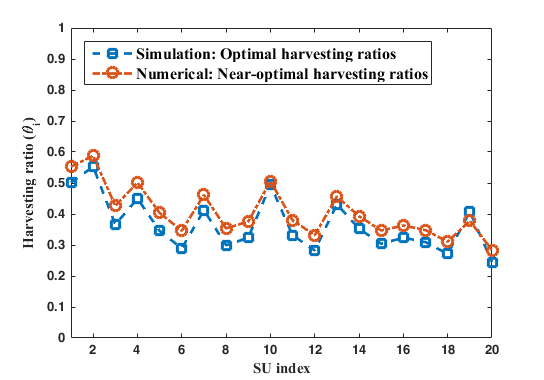}
	\caption{The harvesting ratios for 20 SUs. The energy harvesting rates for SUs are set to $\chi=5$ J/s.}
	\label{fig:HarvestingRatiosSimAndNumerical1}
\end{figure}

After sub-channel allocation process, the SUs receive their sub-channels and the general optimization problem is reduced to a convex NLP. In this part the channel gains are modeled as described in simulation setup. Fig. \ref{fig:HarvestingRatiosSimAndNumerical1}, compares the optimal harvesting ratios depicted by orange circles with our numerical results proposed in Theorem~\ref{LambertTheorem}. There are 20 SUs whose energy harvesting rates are uniformly set to $\chi=5$ J/s. Each SU receives fixed number ($f$) of sub-channels and the interference thresholds of PUs are $5\times10^{-13}$W. It is obvious that our proposed numerical results are capable to obtain more than 95$\%$ of the optimal harvesting ratios for all SUs, which means the performance gap between our proposal and the optimal solution is negligible.

\begin{figure}[t]
	\centering
	\includegraphics[width=9cm,height=7cm]{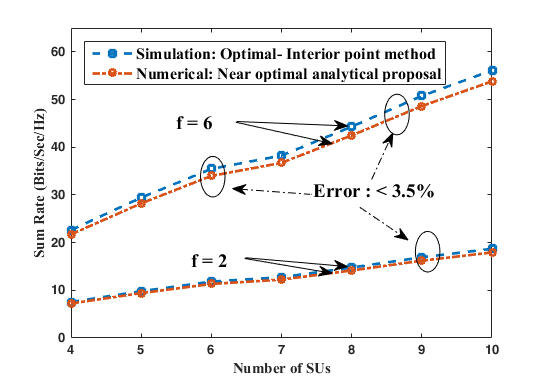}
	\caption{The total sum rate versus number of SUs for a fixed number of allocated sub-channels. Energy harvesting rate is set at $\chi=5$ J/s.}
	\label{fig:SumRateSimAndNumericalAfterSubchannel1}
\end{figure}

Fig. \ref{fig:SumRateSimAndNumericalAfterSubchannel1} shows the sum capacity as a function of the number of SUs, which varies from 4 to 10. The energy harvesting rate is set to $\chi=5$ J/s. Note that each SU receives a fixed number of $f$ sub-channels. We can observe from Fig. \ref{fig:SumRateSimAndNumericalAfterSubchannel1} that the sum rate of all SUs increases when the number of SUs grows from 4 to 10. Two scenarios, each SU getting $f=2$ and $f=6$ sub-channels, respectively, are considered. Since \textbf{P3} is a convex optimization problem, optimal solutions can be obtained by interior point methods. Note that the near optimal theoretical results are within 3.5$\%$ away from the optimal solutions for all cases.

\subsection{Sum Rate versus Rate Constraints}

\begin{figure}[t]
	\centering
	\includegraphics[width=9cm,height=7cm]{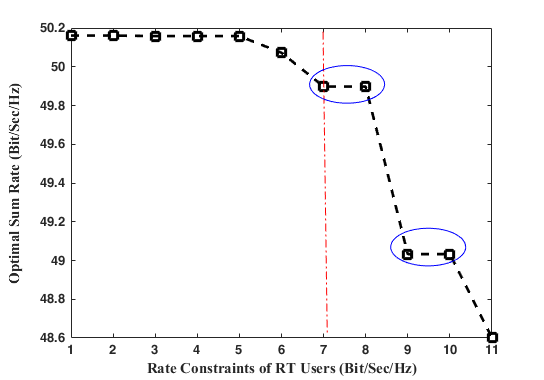}
	\caption{The sum rate versus RT rate constraints ($R^{req}$). ${{\mathcal{K}_R}}=4$ SUs and 16 available sub-channels.}
	\label{fig:SumRateVsRateConstraints2}
\end{figure}

We depict the sum rate of SUs versus different values of rate constraint of RT SUs in Fig. \ref{fig:SumRateVsRateConstraints2}, in which ${{\mathcal{K}_R}}=4$ RT SUs and the available sub-channels are 16 with different channel gain for each SU. The channel gains are detailed in Subsection \ref{sec:SimulSetup}. Each sub-channel has a bandwidth of 62.5 KHz and the harvesting rate is assumed to be $\chi=5$ J/s for all SUs. The required rates of RT users vary from $R_i^{req}= 1$ to $11$ bps/Hz. The time slot duration is considered $T=1$ ms and the sensing time is $\tau_s= 10$ $\mu$s for all SUs. As shown in Fig. \ref{fig:SumRateVsRateConstraints2}, the highest sum rate is achieved for the lowest rate constraint ($R_i^{req}\le 4$ bps/Hz) due to the fact that the optimizer has more freedom to allocate sub-channels to SUs. By increasing the rate constraint from 4 bps/Hz, the sum rate witnesses a slight decrease. Beyond the rate constraint of 7 bps/Hz which is shown by a red line in the figure, the sum rate experiences a sharp decrease. This stems from the fact that the optimizer has less freedom to allocate sub-channels to SUs and thus the optimal sum rate is greatly reduced.
\subsection{System Performance versus Interference Threshold}

\begin{figure}[t]
	\centering
	\includegraphics[width=9cm,height=7cm]{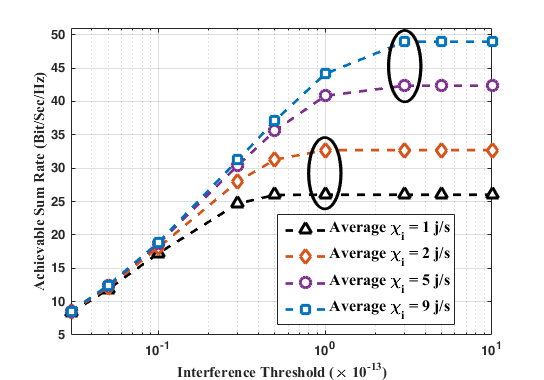}
	\caption{The sum rate versus different PUs' interference thresholds (in Watt). ${{\mathcal{K}_R}}=4$ SUs and 16 available sub-channels.}
	\label{fig:SumRateInterferenceCompare}
\end{figure}

The sum rate of all SUs versus interference thresholds of PUs are illustrated in Fig. \ref{fig:SumRateInterferenceCompare}. Four SUs occupy 16 available OFDM sub-channels. The channel gains provided in Subsection \ref{sec:SimulSetup} are adopted and each SU has a different channel gain. We assume that all PUs have identical interference threshold, which varies between -90 dbm to -110 dbm. We assume that the rate constraints are $R_i^{req}$= 5 bps/Hz for each SU. As can be seen from Fig. \ref{fig:SumRateInterferenceCompare}, the sum rate increases with the growth of the interference threshold. For lower interference thresholds, the SUs' transmission powers are limited and cannot be increased to its maximum amount, thus resulting in lower sum rate performance. However, the sum rate cannot be increased beyond a certain point as the transmission power, which depends on the harvesting ratio, has reached its maximum. Therefore, the sum rate does not increase after the black ellipsoids shown in the figure for different harvesting ratios.

\begin{figure}[t]
	\centering
	\includegraphics[width=9cm,height=7cm]{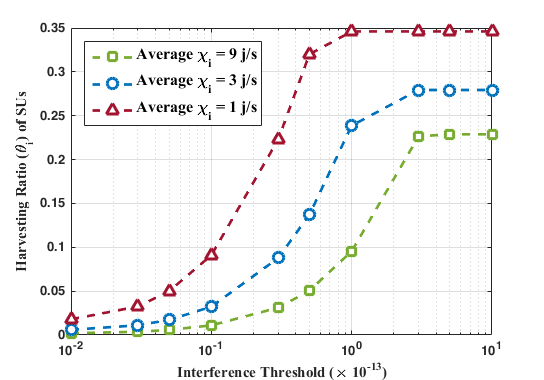}
	\caption{The harvesting ratios versus interference threshold (in Watt). The average harvesting rates ($\chi_i$) are 1, 3, and 9 J/s.}
	\label{fig:ThetaInterferenceCompare}
\end{figure}

We also verify the effect of various PUs' interference thresholds for the achievable harvesting ratios of SUs. Fig. \ref{fig:ThetaInterferenceCompare} illustrates the harvesting ratios versus different PUs' interference thresholds for three cases, where the average harvesting rates are $\chi$ = 1, 3 and 9 J/s, respectively. When the interference threshold is relatively small, SU's power and the sub-channels are interference limited. Thus, the harvesting ratios decrease because the lower interference thresholds require SUs to transmit their data with limited transmission power. Therefore, the harvesting ratios are smaller for lower interference thresholds. However, when the interference threshold increases from $10^{-15}$ W to $3\times10^{-13}$ W, the harvesting ratio grows exponentially in order to extract more energy for data transmission. While the harvesting ratio increases sharply from $I_m^{th}=10^{-15}$ W to $I_m^{th}=3\times10^{-13}$ W, it remains almost constant for higher PUs' interference thresholds. This phenomenon stems from the trade off between having more harvesting time for energy harvesting, and less time for data transmission. In other words, for higher interference thresholds, the harvesting ratio cannot converge to $\theta=1$ because the higher value of $\theta$ implies less time for data transmission. 

\section{Conclusion}\label{sec:conclusion}
In this paper, we have studied the joint resource allocation and structure optimization in multiuser OFDM-based HCRN. We have formulated the general problem of maximizing the sum rate of SUs in green powered OFDM based HCRN under consideration of some practical limitations such as various traffic demands of SUs, interference constraints and imperfect spectrum sensing. We have considered some practical limitations such as various traffic demands of SUs, interference constraint and imperfect spectrum sensing. Then, the general problem of joint resource allocation and structure optimization is formulated as an MINLP task. Since the general problem is NP-hard and intractable, we have tackled the problem in two steps. First, we have proposed a sub-channel allocation scheme based on a factor called Energy Figure of Merit to approximately satisfy SUs' rate requirements and remove the integer constraints. Second, we have proved that the general optimization problem is reduced to a nonlinear convex optimization task. Since the reduced optimization problem cannot achieve exact closed-form solutions, we have thus proposed near optimal closed-form solutions by applying Lambert-W function. We have also exploited the iterative gradient method based on Lagrangian dual decomposition to achieve near optimal solutions. The optimum fractions of the time slot that each SU can harvest energy from the environment are finally obtained.

\appendices
\section{Proof of Lemma 1}
Consider the rate formula in our problem:
\begin{equation}
\left( {1 - {\theta _i} - \frac{{{\tau _{{s_i}}}}}{T}} \right)lo{g_2}\left( {1 + {H_{i,j}}\frac{{{\chi _i}{\theta _i}T - {\epsilon _{{s_i}}}}}{{T - {\theta _i}T - {\tau _{{s_i}}}}}} \right).
\end{equation}
Since the objective function is a sum of rates, if we prove that the rate formula is convex, then the whole objective function becomes a convex problem. Thus,
\begin{equation}
\begin{array}{l}
\begin{split}
&\left( {1 - {\theta _i} - \frac{{{\tau _{{s_i}}}}}{T}} \right)lo{g_2}\left( {1 + {H_{i,j}}\frac{{{\chi _i}{\theta _i}T - {\varepsilon _{{s_i}}}}}{{T - {\theta _i}T - {\tau _{{s_i}}}}}} \right) = \\
&\underbrace {\left( {1 - {\theta _i} - \frac{{{\tau _{{s_i}}}}}{T}} \right){{\log }_2}(T - {\theta _i}T - {\tau _{{s_i}}} + {H_{i,j}}{\chi _i}{\theta _i}T - {H_{i,j}}{\epsilon _{{s_i}}})}_{\mathbf{A}}\\
&\underbrace { - \left( {1 - {\theta _i} - \frac{{{\tau _{{s_i}}}}}{T}} \right){{\log }_2}(T - {\theta _i}T - {\tau _{{s_i}}})}_{\mathbf{B}}.
\end{split}
\end{array}
\end{equation}
Consider the second part (B):\\
\begin{equation}
\begin{split}
{\bf{B}} &=  - {\left( {1 - {\theta _i} - \frac{{{\tau _{{s_i}}}}}{T}} \right)}{\log _2}(T - {\theta_i}T - {\tau _{{s_i}}})\\
&=  - \frac{1}{T}(T - {\theta_i}T - {\tau _{{s_i}}}){\log _2}(T - {\theta_i}T - {\tau _{{s_i}}})\\
&= - \frac{1}{T}Z{\log _2}(Z),
\end{split}
\end{equation}
where Z is equal to \textbf{C2} in \textbf{P1} and it is always greater than zero ($Z>0$). Thus, the second part is similar to the famous form of concave functions ($-Xlog(X)$). Thus, this part is proved to be concave.\\ 
Now, consider the first part (A):\\
\begin{equation}
{\bf{A}} = {\left( {1 - {\theta _i} - \frac{{{\tau _{{s_i}}}}}{T}} \right)}{\log _2}(T - {\theta _i}T - {\tau _{{s_i}}} + {H_{i,j}}{\chi_i}{\theta_i}T - {H_{i,j}}{\epsilon_{{s_i}}})
\end{equation}
Then, we take the second derivative (Hessian) with respect to $\theta_i$:\\
\begin{equation}
\begin{gathered}
\left[ {\frac{{\left( {{\tau _{{s_i}}} - T + 2{H_{i,j}}{\epsilon _{{s_i}}} + {\theta _i}T} \right) \times T({H_{i,j}}{\chi _i} - 1)}}{{{{({\tau _{{s_i}}} - T + {H_{i,j}}{\epsilon _{{s_i}}} + {\theta _i}T - {H_{i,j}}{\chi _i}{\theta _i}T)}^2}}}} \right. +  \hfill \\
\left. {\frac{{\left( { - {H_{i,j}}{\chi _i}T + {H_{i,j}}{\chi _i}{\tau _{{s_i}}} - {H_{i,j}}{\chi _i}{\theta _i}T} \right) \times T({H_{i,j}}{\chi _i} - 1)}}{{{{({\tau _{{s_i}}} - T + {H_{i,j}}{\epsilon _{{s_i}}} + {\theta _i}T - {H_{i,j}}{\chi _i}{\theta _i}T)}^2}}}} \right]. \hfill \\ 
\end{gathered}
\end{equation}
We need to prove that the second derivative is less than zero. The denominator is always positive:
\begin{equation}
{({\tau _{{s_i}}} - T + {H_{i,j}}{\epsilon_{{s_i}}} + {\theta_i}T - {H_{i,j}}{\chi_i}{\theta_i}T)^2} > 0
\end{equation}
Thus, we consider the nominator
\begin{equation}
\begin{split}
&{\tau _{{s_i}}} - T + 2{H_{i,j}}{\epsilon_{{s_i}}} + {\theta_i}T - {H_{i,j}}{\chi_i}T + {H_{i,j}}{\chi_i}{\tau _{{s_i}}} -\\
&{H_{i,j}}{\chi_i}{\theta_i}T.
\end{split}
\end{equation}
By adding and subtracting ${H_{i,j}}{\chi_i}{\theta_i}T$, we have
\begin{equation} \label{SSS}
\begin{split}
&({\tau _{{s_i}}} - T + {\theta_i}T) + {H_{i,j}}( - {\chi_i}{\theta_i}T + {\epsilon_{{s_i}}}) + {H_{i,j}}{\epsilon_{{s_i}}}\\
&- {H_{i,j}}{\chi_i}({\theta_i} + 1 - {\theta_i})T + {H_{i,j}}{\chi_i}{\tau _{{s_i}}} - {H_{i,j}}{\chi_i}{\theta_i}T.
\end{split}
\end{equation}
Further rearranging the terms proves that the nominator is negative.
\begin{equation}
\begin{split}
&\underbrace {({\tau _{{s_i}}} - T + {\theta_i}T)}_{ < 0} + {H_{i,j}}\underbrace {( - {\chi_i}{\theta_i}T + {\epsilon_{{s_i}}})}_{ < 0} +\\& {H_{i,j}}\underbrace {( - {\chi_i}{\theta_i}T + {\epsilon_{{s_i}}})}_{ < 0} + {H_{i,j}}{\chi_i}\underbrace {({\tau _{{s_i}}} - T + {\theta_i}T)}_{ < 0} < 0.
\end{split}
\end{equation}
Thus, $T({H_{i,j}}{\chi_i} - 1) > 0$ (i.e., ${H_{i,j}}{\chi_i} > 1$) must be hold such that the whole objective function becomes convex for minimization (concave for maximization). Then, the convex problem can be solved by standard convex optimization techniques.

\section{Proof of Lemma 2}

To prove \textbf{P1} to be convex MINLP, one should consider convexity of constraints functions $\forall i \in \mathcal{K}$, $0 < {\theta _i} < 1$ \cite{00233}. Thus, we consider the constraints \textbf{C3}, \textbf{C4}, and \textbf{C5}. The constraint function of \textbf{C3} has the function of $\frac{{{\chi _i}{\theta _i}T - {\epsilon_{{s_i}}}}}{{T - {\theta _i}T - {\tau _{{s_i}}}}}$ and its derivative is 
\begin{equation}
\frac{{{\chi _i}T\left( {T - {\theta _i}T - {\tau _{{s_i}}}} \right) + T\left( {{\chi _i}{\theta _i}T - {\epsilon_{{s_i}}}} \right)}}{{{{\left( {T - {\theta _i}T - {\tau _{{s_i}}}} \right)}^2}}},
\end{equation}
where the numerator can be simplified, and the derivative function is always positive $\frac{{T\left( {{\chi _i}\left( {T - {\tau _{{s_i}}}} \right) - {\epsilon_{{s_i}}}} \right)}}{{{{\left( {T - {\theta _i}T - {\tau _{{s_i}}}} \right)}^2}}} > 0$ over the $\theta_i \in (0,1)$. Then, the second derivative with respect to $\theta_{i}$ is 
\begin{equation}
\frac{{2{T^2}\left( {{\theta _i}^2T - \left( {T - {\tau _{{s_i}}}} \right)} \right)\left( {{\chi _i}\left( {T - {\tau _{{s_i}}}} \right) - {_{{s_i}}}} \right)}}{{{{\left( {T - {\theta _i}T - {\tau _{{s_i}}}} \right)}^4}}},
\end{equation}
where the denominator is always positive; however, the term of ${\left( {{\theta _i}^2T - \left( {T - {\tau _{{s_i}}}} \right)} \right)}$ in the numerator is always negative since $T-\tau_{{s_i}}$ is greater than ${{\theta _i}^2T}$. Thus, the second derivative is negative and the constraint \textbf{C3} is a concave function for maximization problem (convex for standard minimization). Meanwhile, the convexity of constraints \textbf{C4} and \textbf{C5} can be proved similarly to the proof of Lemma 1.

\section{Proof of Proposition 1}

In order to avoid convergence to a non-stationary point, some of step sizes should be infinite. Meanwhile, with respect to the iteration index, the step sizes ${\alpha ^t}, {\beta ^t}, {\pi ^t}, {\psi ^t},$ and ${\eta ^t}$ tend to zero. Thus, the conditions comply the convergence of dual variables to their corresponding dual optimal solutions \cite{Bertsekas,Azimi2016}.

\section{Proof of Lemma 3}

To prove this Lemma, we need to write the equation ($x = \frac{b}{{\mathcal{W}\left( {\frac{b}{e^{a}}} \right)}}$) in the form of the \textit{Lambert} function. Thus, the solution can be derived as follows:
\begin{equation}
\ln \left( x \right) = a + \frac{b}{x}.
\end{equation}
Taking the exponential power from both sides results in
\begin{equation} \label{tpowerExp}
x = {e^{\left( {a + \frac{b}{x}} \right)}}.
\end{equation}
Let $y = \frac{b}{x}$. Then, Eq.~(\ref{tpowerExp}) can be expressed as
\begin{equation} \label{exponentialform}
y{e^y} = \frac{b}{e^{a}}.
\end{equation}
The \textit{Lambert} $\mathcal{W}$ function can now be applied, resulting in
\begin{equation} \label{YLambertEq45}
y = \mathcal{W}\left( {\frac{b}{e^{a}}} \right).
\end{equation}
Finally, substituting $x = \frac{b}{y}$ into Eq. (\ref{YLambertEq45}) results in Lemma 3.

\bibliography{ArxiveJointSpectrumAllocationandStructureOptimizationinGreenPoweredHCRN}
\bibliographystyle{IEEEtr}

\end{spacing}

\end{document}